\documentclass[12pt]{article}

\RequirePackage[OT1]{fontenc}
\usepackage{natbib}
\RequirePackage[colorlinks,citecolor=blue,urlcolor=blue]{hyperref}
\usepackage[english]{babel}
\usepackage{amsthm}
\usepackage{amsmath}
\usepackage{amsfonts}
\usepackage{amssymb}
\usepackage{graphicx}
\usepackage{enumerate}
\usepackage{color}
\usepackage{array}
\usepackage{psfrag,epsf}
\usepackage{url} 
\usepackage[export]{adjustbox}
\usepackage{bm}
\usepackage{titling}
\usepackage{algorithm}
\usepackage{algorithmic}

\def\frac#1#2{{\textstyle{#1\over#2}}}

\DeclareSymbolFont{AMSb}{U}{msb}{m}{n}
\DeclareMathSymbol{\Natural}{\mathbin}{AMSb}{"4E}
\DeclareMathSymbol{\Integer}{\mathbin}{AMSb}{"5A}
\DeclareMathSymbol{\Real}{\mathbin}{AMSb}{"52}
\DeclareMathSymbol{\Rational}{\mathbin}{AMSb}{"51}
\DeclareMathSymbol{\Imaginary}{\mathbin}{AMSb}{"49}
\DeclareMathSymbol{\Complex}{\mathbin}{AMSb}{"43} 
\DeclareMathSymbol{\Disk}{\mathbin}{AMSb}{"44} 
\def\bi{\begin{itemize}}
\def\ei{\end{itemize}}
\def\bd{\begin{description}}
\def\ed{\end{description}}
\def\ben{\begin{enumerate}}
\def\een{\end{enumerate}}
\def\bv{\begin{verbatim}}
\def\ev{\end{verbatim}}

\def\bar#1{{\overline{#1}}}
\def\hat#1{{\widehat{#1}}}

\def\2to{{\ {\buildrel 2\over \longrightarrow}\ }}

\def\dotsim{{\ {\buildrel \cdot\over \sim}\ }}
\def\I1ton{{$I_1,\ldots,I_n$}}
\def\X1ton{{$X_1,\ldots,X_n$}}
\def\Y1ton{{$Y_1,\ldots,Y_n$}}
\def\Z1ton{{$Z_1,\ldots,Z_n$}}
\def\R1ton{{$R_1,\ldots,R_n$}}
\def\e1ton{{$e_1,\ldots,e_n$}}
\def\t1ton{{$t_1,\ldots,t_n$}}
\def\x1ton{{$x_1,\ldots,x_n$}}
\def\y1ton{{$y_1,\ldots,y_n$}}
\def\z1ton{{$z_1,\ldots,z_n$}}

\def\code#1{\texttt{#1}}

\begin{document}
\thispagestyle{empty}
\baselineskip=28pt
\vskip 5mm
\begin{center} {\LARGE{\bf Estimating high-resolution Red Sea surface temperature hotspots, using a low-rank semiparametric spatial model}}
\end{center}


\baselineskip=12pt
\vskip 5mm

\begin{center}
\large
Arnab Hazra$^1$ and Rapha{\"e}l Huser$^1$
\end{center}

\footnotetext[1]{
\baselineskip=10pt Computer, Electrical and Mathematical Sciences and Engineering (CEMSE) Division, King Abdullah University of Science and Technology (KAUST), Thuwal 23955-6900, Saudi Arabia. \\ E-mails: arnab.hazra@kaust.edu.sa; raphael.huser@kaust.edu.sa}

\baselineskip=17pt
\vskip 4mm
\centerline{\today}
\vskip 6mm

\begin{center}
{\large{\bf Abstract}}
\end{center}

In this work, we estimate extreme sea surface temperature (SST) hotspots, i.e., high threshold exceedance regions, for the Red Sea, a vital region of high biodiversity. We analyze high-resolution satellite-derived SST data comprising daily measurements at 16703 grid cells across the Red Sea over the period 1985--2015. We propose a semiparametric Bayesian spatial mixed-effects linear model with a flexible mean structure to capture spatially-varying trend and seasonality, while the residual spatial variability is modeled through a Dirichlet process mixture (DPM) of low-rank spatial Student-$t$ processes (LTPs). By specifying cluster-specific parameters for each LTP mixture component, the bulk of the SST residuals influence tail inference and hotspot estimation only moderately. Our proposed model has a nonstationary mean, covariance and tail dependence, and posterior inference can be drawn efficiently through Gibbs sampling. In our application, we show that the proposed method outperforms some natural parametric and semiparametric alternatives. { Moreover, we show how hotspots can be identified and we estimate extreme SST hotspots for the whole Red Sea, projected until the year 2100, based on the Representative Concentration Pathways 4.5 and 8.5.} The estimated 95\% credible region for joint high threshold exceedances include large areas covering major endangered coral reefs in the southern Red Sea.

\baselineskip=16pt

\par\vfill\noindent

\pagenumbering{arabic}
\baselineskip=25pt

\newpage

\section{Introduction}
\label{sec:intro}



Sea surface temperature (SST) has an immense environmental and ecological impact on marine life and ecosystems, e.g., affecting the survival of endangered animal species including corals \citep{reaser2000coral,berumen2013status,lewandowska2014effects}, and also has an important economic impact for neighboring countries, which depend on it for their local fisheries and tourism. 
Hence, the identification of the regions within the Red Sea where SST may exceed high thresholds 
is a vital concern and this motivates a proper statistical analysis of (present and future) extreme hotspots from a high resolution spatiotemporal SST dataset. Operational Sea Surface Temperature and Sea Ice Analysis (OSTIA) produces satellite-derived daily SST data at $0.05^\circ \times 0.05^\circ$ resolution \citep{donlon2012operational}. Over the whole Red Sea, daily SST data are available at 16703 grid cells between 1985--2015 and we consider these data for estimating extreme hotspots.

The most common model in spatial geostatistics is the Gaussian process (GP) due to its appealing theoretical and computational properties \citep{gelfand2016spatial}. However, fitting an ordinary GP model involves computing the determinant and the inverse of the spatial covariance matrix, which is excessively prohibitive in dimensions as high as the Red Sea SST data (here, available at 16703 grid cells). A variety of methods have been proposed to tackle this problem. These include approaches based on kernel convolutions \citep{higdon2002space}, low-rank methods using basis functions \citep{wikle1999dimension}, the predictive process \citep{banerjee2008gaussian}, approximations of the likelihood in the spectral domain \citep{stein1999kriging, fuentes2007approximate} or by a product of appropriate conditional distributions \citep{vecchia1988estimation, stein2004approximating}, covariance tapering \citep{furrer2006covariance, anderes2013nonstationary} and Markov random fields \citep{rue2005gaussian, rue2009approximate}; {see \citet{Heaton.etal:2019} for a comparative overview of (some of) these methods.} Irrespective of being an ordinary GP or a low-rank GP (LGP) model, the marginal normal density functions are thin-tailed and hence they can heavily underestimate the probabilities of extreme events. Additionally, the tails of multivariate normal distributions lead to independent extremes {asymptotically} except in the trivial case of perfect dependence which can result in disastrous underestimation of the  simultaneous occurrence probabilities of extreme events \citep{davison2013geostatistics}. Hence, both GPs and LGPs have been criticized when the main interest lies in the tail behavior. Relaxing the parametric GP assumption, \cite{gelfand2005bayesian} propose a flexible nonparametric Bayesian model based on a Dirichlet process mixture (DPM) of spatial GPs in the context of geostatistical analysis; however, \cite{hazra2018semiparametric} showed that the {joint tail of} a finite mixture of GPs also leads to independent extremes. There are more flexible nonparametric spatial models available in the geostatistics literature (see, e.g., \citealp{duan2007generalized}), but these are { often} not computationally suitable for large spatial datasets.

While GPs and related processes are typically used to describe the bulk behavior, models stemming from extreme-value theory are designed to accurately describe the tail behavior. The classical modeling of spatial extremes usually relies on site-wise block maxima or peaks over some high threshold \citep{smith1990max, davison2012statistical,davison2015statistics, davison2019spatial,Huser.Wadsworth:2020}. They can be divided into three main categories: (asymptotic) max-stable and Pareto processes \citep{smith1990max, padoan2010likelihood, davison2012statistical, reich2012hierarchical, opitz2013extremal, thibaud2015efficient, de2018high}, latent variable models \citep{sang2009hierarchical, sang2010continuous, cooley2010spatial, opitz2018inla, CastroCamilo.etal:2019} and  sub-asymptotic models \citep{huser2017bridging, morris2017space,  hazra2019multivariate, huser2019modeling}. Max-stable and Pareto processes are asymptotically justified models for spatial extremes but likelihood computations are usually challenging even for low or moderate spatial dimensions (see, e.g., \citealp{Castruccio.etal:2016} and \citealp{Huser.etal:2019}). An exception is the max-stable model of \cite{reich2012hierarchical}, which is computationally tractable for higher spatial dimensions; see also \citet{Bopp.etal:2020}. {However, this max-stable model has been criticized for its lack of flexibility in a variety of applications.} Analogously, \cite{de2018high} show how Pareto processes can be efficiently fitted to high-dimensional peaks-over-thresholds using proper scoring rules, but their approach is limited to a few thousand sites and cannot easily be accommodated to the Bayesian setting. Alternatively, \cite{morris2017space} propose a Bayesian model for high threshold exceedances based on the spatial skew-$t$ process, which lacks the strong asymptotic characterization of max-stable and Pareto processes but benefits from exceptionally faster computation. Instead of considering only block maxima or peaks-over-thresholds as in the above-mentioned approaches, \cite{hazra2018semiparametric} consider a DPM of spatial skew-$t$ processes where the extremes are selected probabilistically through the Dirichlet process prior; hence, this modeling approach does not require any arbitrary high thresholding but assumes that identically distributed temporal replicates are available. More recently, \cite{bopp2020projecting} have proposed a Bayesian analogue model where the intensity of flood-inducing precipitation is modeled using a mixture of Student's $t$ processes. However, these approaches are applicable only for relatively low spatial dimensions and they are not directly applicable for large spatiotemporal datasets with trend and seasonality, such as our Red Sea SST dataset.


In this paper, we propose a low-rank semiparametric Bayesian spatial mixed-effects linear model, which extends the spatial model of \cite{hazra2018semiparametric}, to handle large, highly nonstationary spatiotemporal datasets. { Following \cite{fix2018comparison}, we assume the mean SST profile to be comprised of a spatially-varying trend that is linearly related to the mean annual sea surface temperature projections for the Red Sea under the Representative Concentration Pathway (RCP) 4.5 and 8.5, adopted by the Intergovernmental Panel on Climate Change (IPCC), where RCP 8.5 corresponds to the pathway with high greenhouse gas emissions, while RCP 4.5 describes a moderate mitigation pathway.} Additionally, we also consider a nonlinear seasonality term (modeled using B-splines) and for computational tractability with high spatial resolution, we consider a low-rank approximation of the spatially-varying coefficients involved within the trend and seasonality components. { While the use of splines implies some degree of regularity, it is justified in our case since SST data vary smoothly over space. Moreover, unlike the max-stable model of \cite{reich2012hierarchical} whose dependence structure is constructed from smooth kernels, we here only consider the mean structure to be comprised of splines, while the dependence structure of residuals has a more flexible stochastic representation.} Specifically, we model the {residual} variability using a DPM of low-rank spatial Student-$t$ processes (LTPs), abbreviated by LTP-DPM in short. The DPM is constructed using a (truncated) stick-breaking prior \citep{sethuraman1994constructive}. The LTP mixture components are constructed by a scalar product of multivariate normal random effects with orthonormal spatial basis functions, and then multiplied by an inverse-gamma random scaling. The random effects are assumed to be independent and identically distributed (iid) across time points. The proposed model has nonstationary mean, covariance and tail dependence structure, {and under a suitable asymptotic regime, its covariance structure spans the class of all covariance structures from squared-integrable stochastic processes with continuous covariance functions.}
We draw posterior inference about the model parameters through an efficient Gibbs sampler.

Beyond modeling the SST data, our ultimate goal is to identify regions ``at risk'', where the SST level might exceed a high threshold at some future time. Similar problems arise in a wide range of scientific disciplines, for example, environmental health monitoring \citep{bolin2015excursion}, brain imaging \citep{mejia2020bayesian}, astrophysics \citep{beaky1992topology} and climatology \citep{furrer2007spatial, french2013spatio}. The easiest and most naive approach for estimating exceedance regions (i.e., hotspots) is to perform site-specific exceedance tests at each grid cell separately (see, e.g., \citealp{eklundh2003vegetation}). {However, if such an approach does not adequately account for multiple testing nor spatial dependence, it will fail to accurately represent the co-occurrence of extreme events.} A better approach { for estimating hotspots} is to set the joint probability of exceeding a high threshold over the \emph{whole} region being equal to some predefined value. In this spirit, a variety of more advanced methods have been proposed for identifying hotspots \citep[see, e.g.,][]{Cressie.etal:2005,Craigmile.etal:2006,french2013spatio,bolin2015excursion}. In particular, \cite{french2013spatio} provide a sampling-based method for constructing confidence regions for Gaussian processes that contain the true exceedance regions with some predefined probability. Here, we develop a similar approach to estimate extreme hotspots by extending the Gaussian-based method of \cite{french2013spatio} and \cite{french2016credible} to the more general framework of our semiparametric LTP-DPM model, which is better suited for capturing the joint tail behavior of complex spatiotemporal processes.

The paper is organized as follows. In \S\ref{data}, we present the Red Sea SST dataset and some exploratory analysis. Our proposed LTP-DPM model and its properties are discussed in \S\ref{methodology}. In \S\ref{computation}, we discuss Bayesian computational details and and the hotspot estimation technique. In \S\ref{application}, we apply the proposed methodology to the Red Sea SST dataset and discuss the results. We conclude with some discussion and perspectives on future research in \S\ref{discussions}.

\section{The Red Sea SST dataset and exploratory analysis}
\label{data}

\begin{figure}[t!]
\begin{center}
    	\adjincludegraphics[height = 0.4\linewidth, width = 0.25\linewidth, trim = {{.12\width} {.0\width} {.27\width} {.0\width}}, clip]{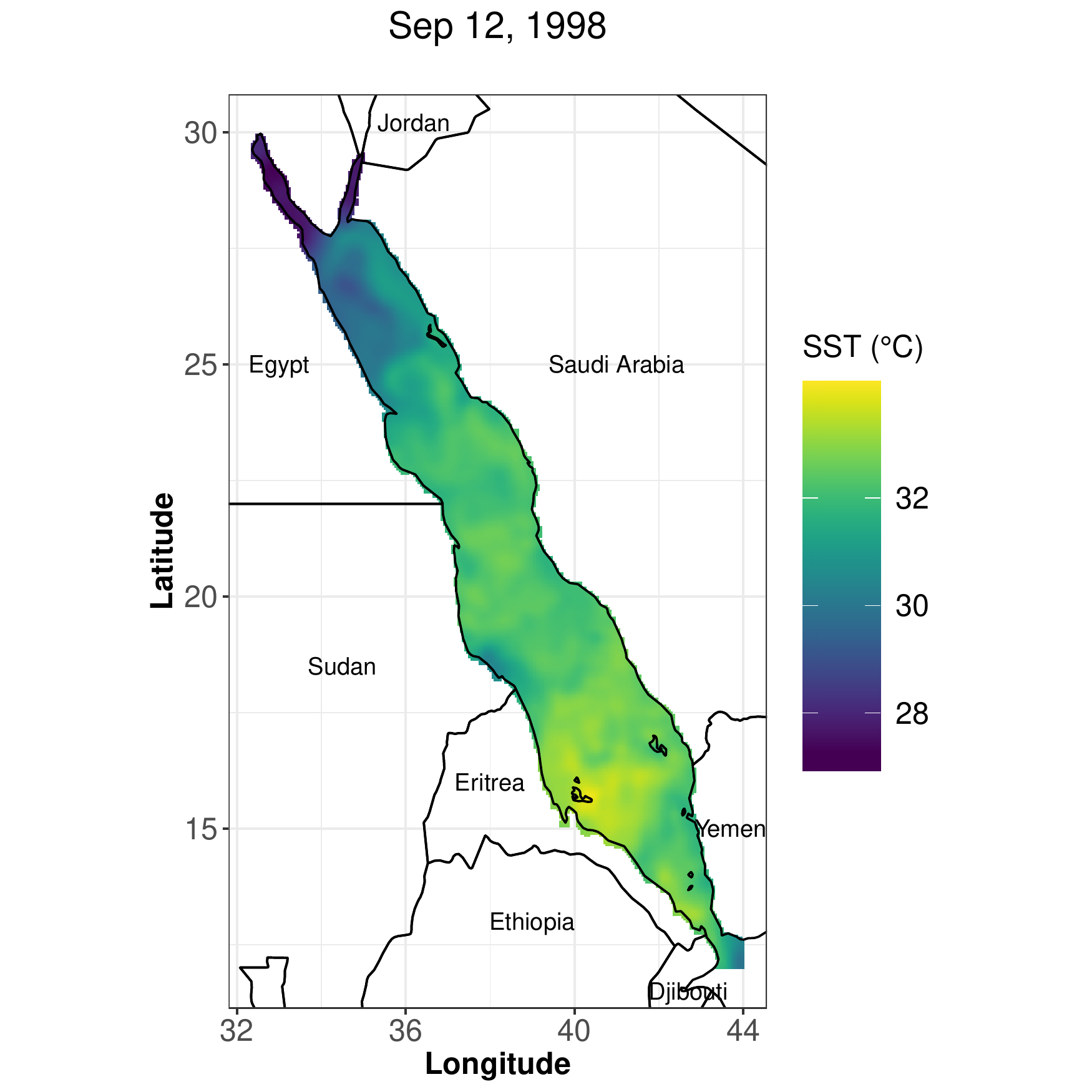}
	\adjincludegraphics[height = 0.4\linewidth, width = 0.25\linewidth, trim = {{.12\width} {.0\width} {.27\width} {.0\width}}, clip]{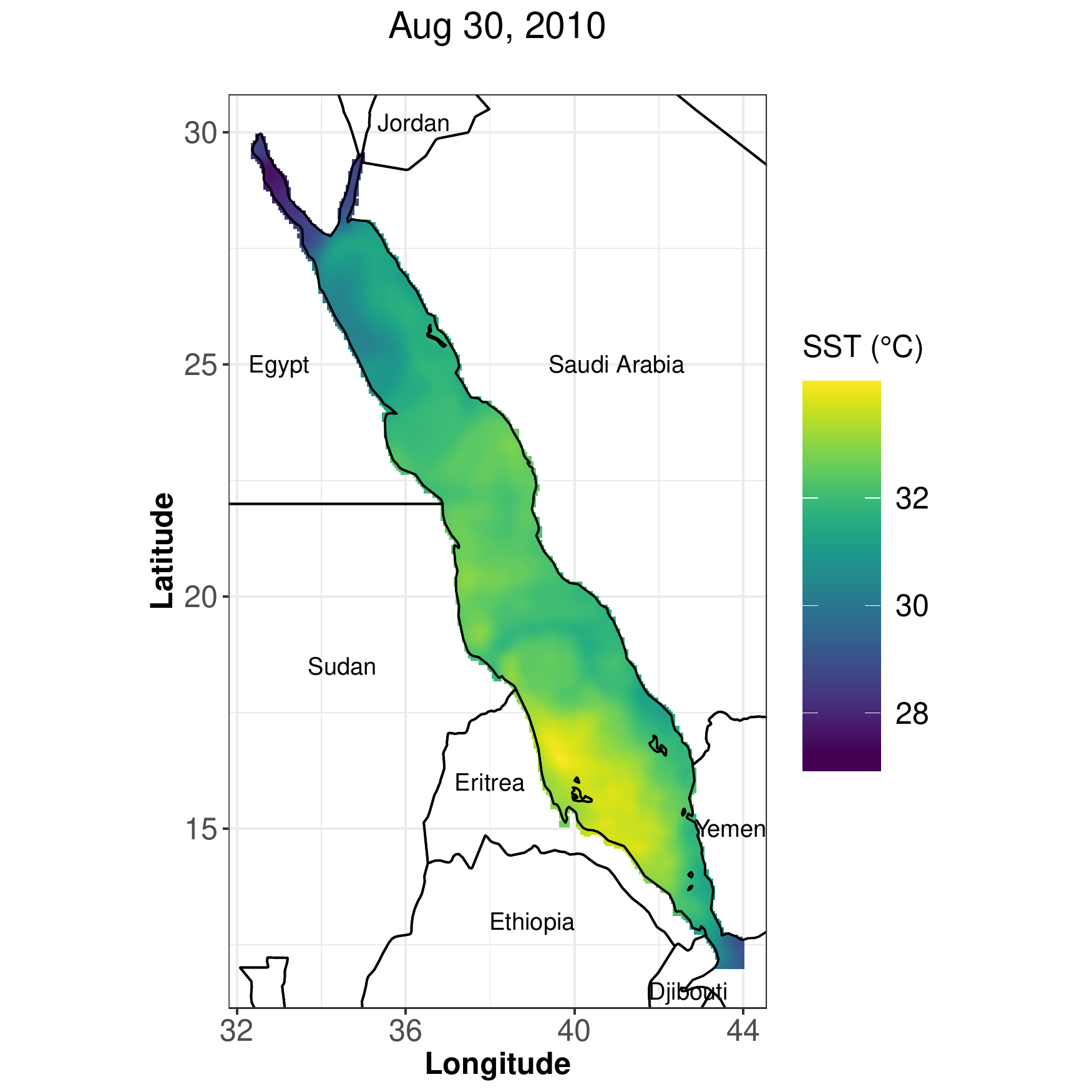}
	\adjincludegraphics[height = 0.4\linewidth, width = 0.36\linewidth, trim = {{.12\width} {.0\width} {.0\width} {.0\width}}, clip]{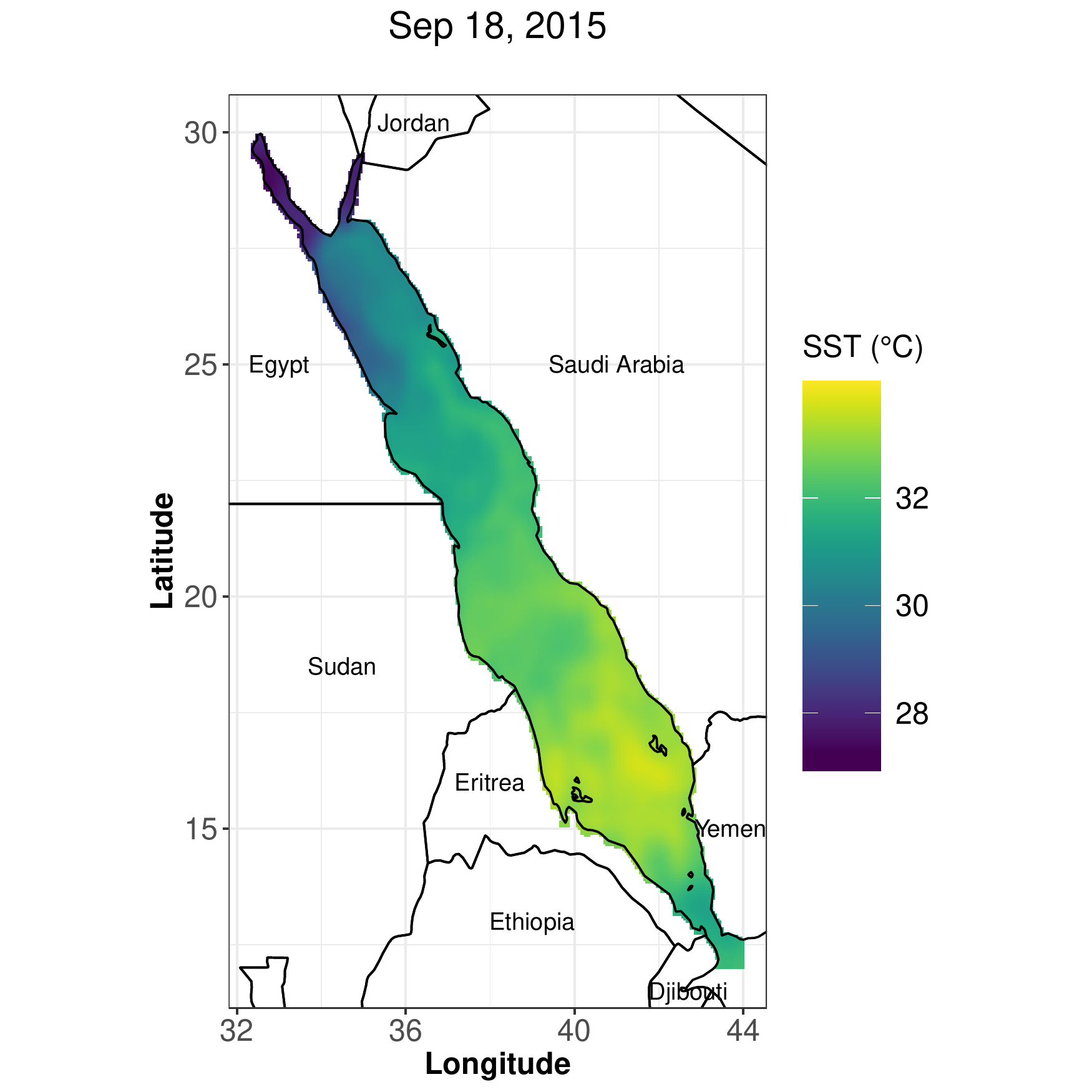}
\end{center}
	\caption{Observed SST profiles across the Red Sea for three extremely hot days: September 12, 1998 (left), August 30, 2010 (middle), and September 18, 2015 (right). All sub-figures are on the same scale.}
	\label{three_hotdays}
\end{figure}

The OSTIA project generates satellite-derived daily SST estimates (free of diurnal variability) at an output grid resolution of $0.05^\circ \times 0.05^\circ$ (about 6 km). This yields 16703 grid cells for the whole Red Sea. The data can be freely obtained from the website \url{http://ghrsst-pp.metoffice.com/pages/latest_analysis/ostia.html}, {and were also (partly) analyzed for the data competition organized for the 11-th International Conference on Extreme-Value Analysis, 2019, though with a different objective; see \citet{Huser:2020}.} Figure \ref{three_hotdays} shows spatial maps of observed SST profiles for three different days with high spatially-averaged SST. For all three days, the SST values are lowest near the Gulfs of Aqaba and Suez in the North, and highest in the southern Red Sea near the coast of Eritrea and the southwest of Saudi Arabia.

Some exploratory analysis (not shown) reveals that daily SST data are highly autocorrelated in time, and that the temporal dependence strength varies strongly over space. It would be extremely statistically and computationally challenging---if possible at all---to flexibly account for spatially-varying autocorrelation in a single fully Bayesian model for a dataset of this size ($16703$ grid cells and $11315$ days). Therefore, for simplicity, we here analyze temporally-thinned data, keeping only one day per week at each grid cell, thus greatly reducing the temporal auto-correlation. {Hence, we obtain seven (correlated) sub-datasets, each comprising 1612 spatial fields (i.e., one per week) that we treat as independent time replicates.} The results are observed to be consistent across the sub-datasets. To concisely summarize the results for all sub-datasets, and to reduce the overall uncertainty in our final estimates, we obtain the results separately for each sub-dataset, check that they are indeed mutually consistent, and then report the averages. 
As our main goal is to draw \emph{spatial} inference, and estimate \emph{spatial} (rather than spatiotemporal) hotspots, this approach is reasonable.



\begin{figure}[t!]
\begin{center}
	\adjincludegraphics[height = 0.4\linewidth, width = 0.23\linewidth, trim = {{.12\width} {.0\width} {.32\width} {.0\width}}, clip]{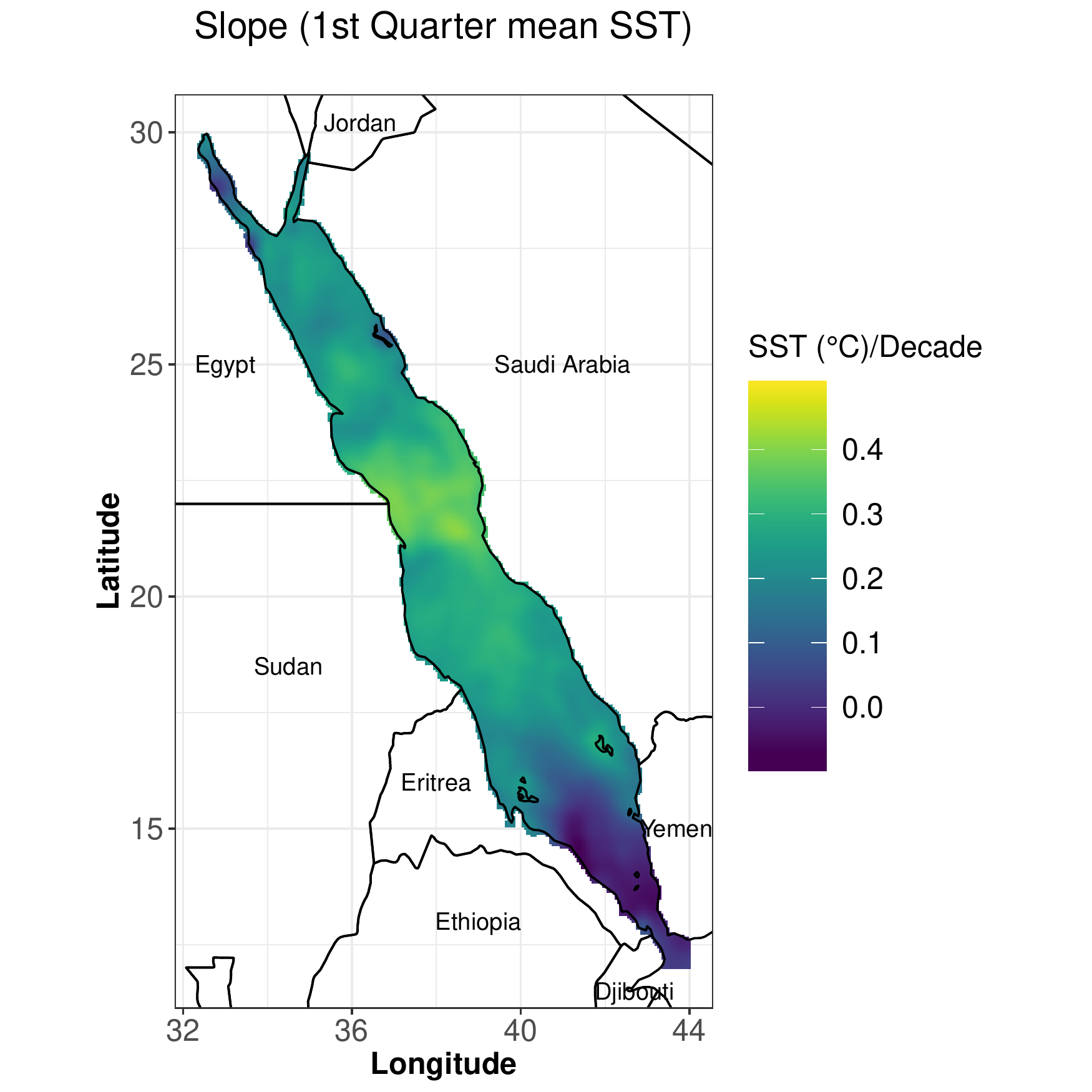}
	\adjincludegraphics[height = 0.4\linewidth, width = 0.23\linewidth, trim = {{.12\width} {.0\width} {.32\width} {.0\width}}, clip]{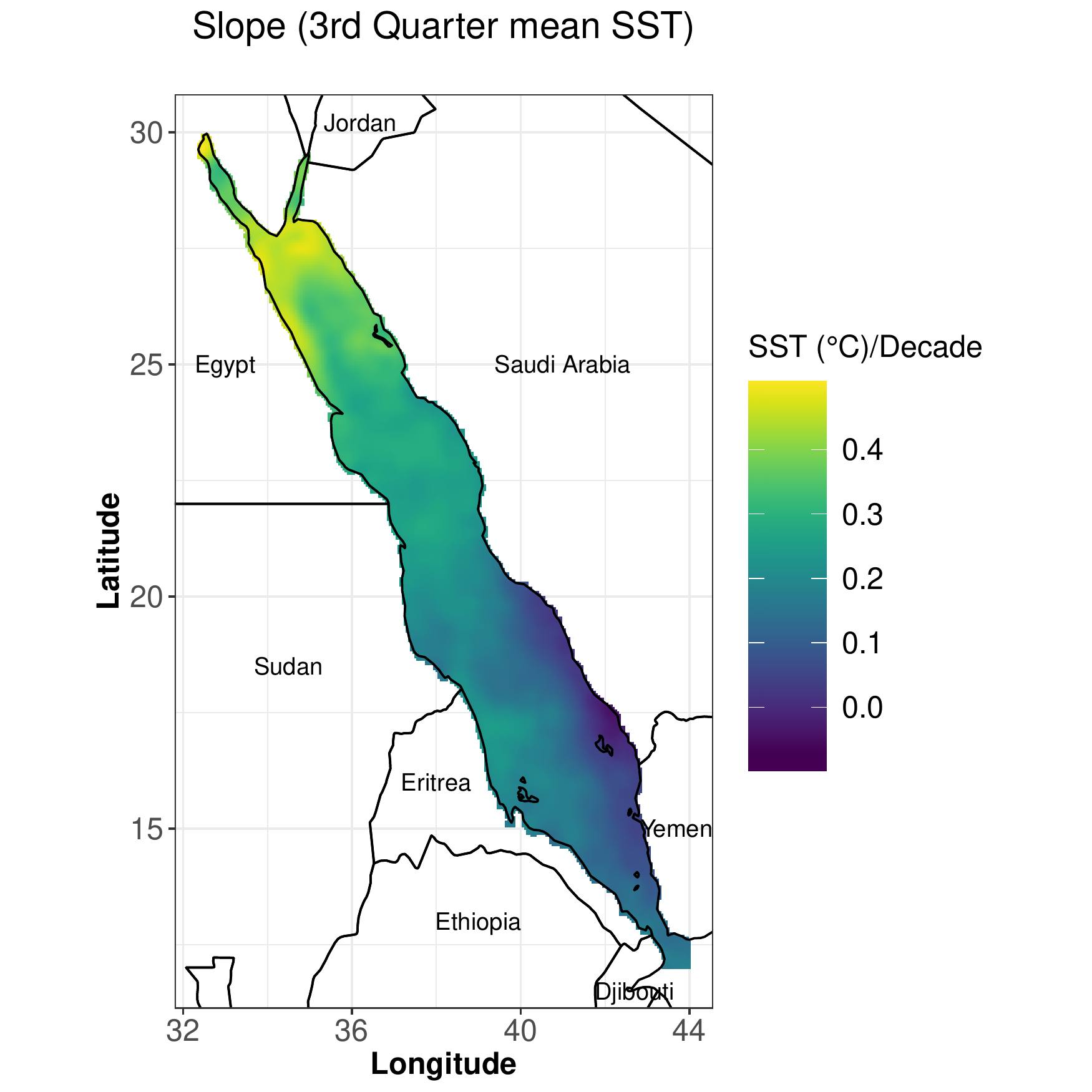}
	\adjincludegraphics[height = 0.4\linewidth, width = 0.32\linewidth, trim = {{.12\width} {.0\width} {.10\width} {.0\width}}, clip]{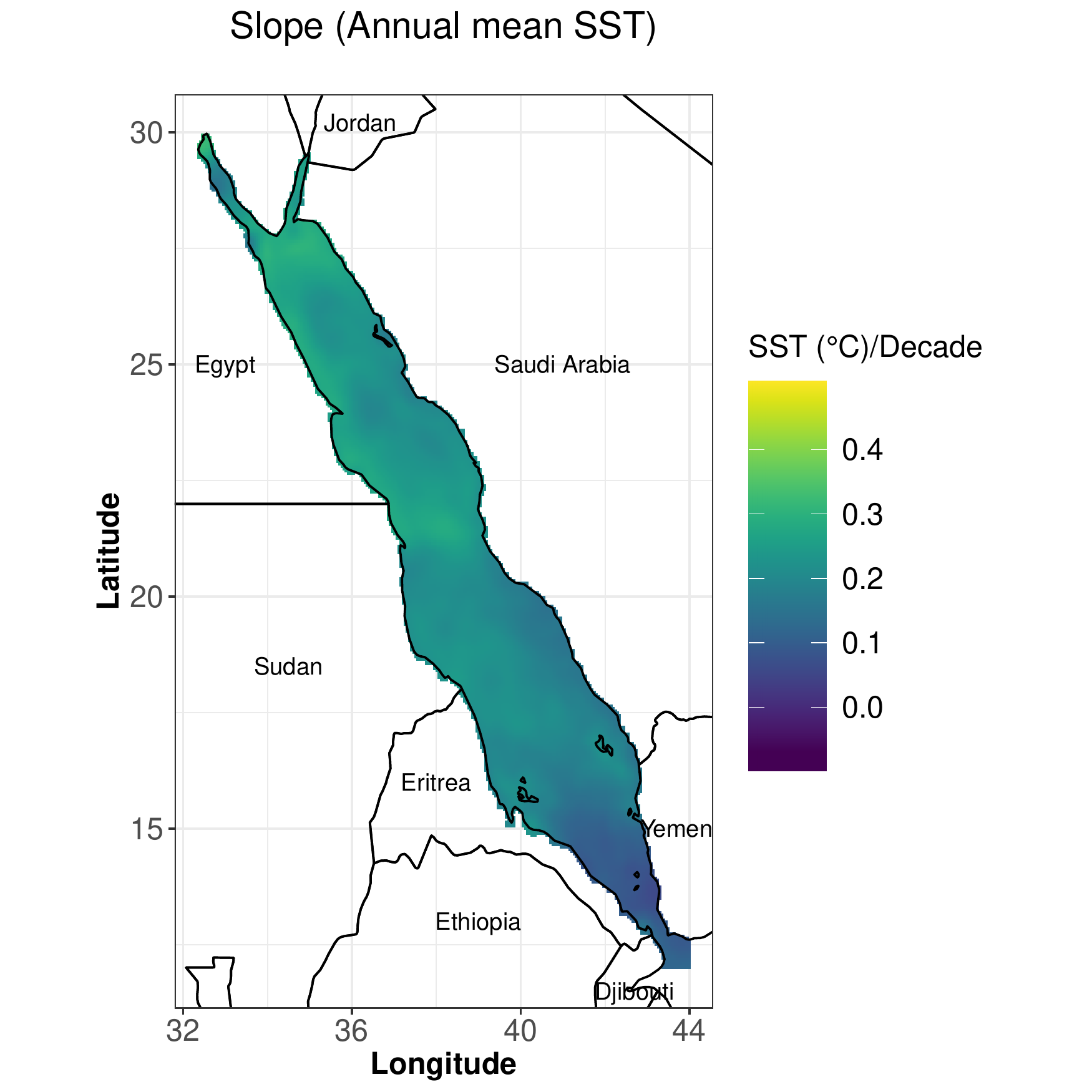}
\end{center}
\caption{Decadal rate of change of the quarterly average SST for the first (left) and the third (middle) quarters and also of the annual average SST (right) over the Red Sea. All sub-figures are on the same scale.}
\label{slopes}
\end{figure}

Considering the spatially-varying nature of SST across the Red Sea and the long data collection period of 31 years from  1985 to 2015, a period that faced global warming, we first conduct a preliminary site-by-site average and trend analysis of the annual mean SST as well as the quarterly mean SST, {by fitting simple linear regression models estimated by least squares}. The mean SST across the Red Sea is lowest during the first quarter and highest during the third quarter. The trend profiles (decadal rate of change) of the first-quarterly, third-quarterly and annual mean SST are displayed in Figure \ref{slopes}. For the first quarterly and annual mean SST, the slopes are the highest near the latitude 22$^\circ$N and are the lowest near the southern end of the Red Sea between Eritrea and Yemen. However, for the third quarterly mean SST, the highest slopes are observed throughout the coast of Egypt in the northwest, and the lowest values are observed near the southwest of Saudi Arabia. Thus, Figure \ref{slopes} explains the need for spatially-, as well as seasonally-, or weekly-varying coefficients for modeling the marginal SST distributions. 


We then study the seasonality profile (averaged across years) at various grid cells. {The results are reported in the Supplementary Material.} Significantly different patterns are observed throughout the Red Sea. The hottest weeks (maximizing the annual weekly average SST) vary mainly between weeks 32 and 35 (above the latitude 20$^\circ$N), and weeks 37 and 42 (below 20$^\circ$N). The observed nonstationarity of SST across weeks explains the need for a flexible modeling of seasonality, e.g., through some linear combination of local basis functions with spatially-varying coefficients. 

{ While the seasonal pattern and the spatial variability of the mean SST profile may be estimated from observed data, future SST projections require the use of an additional covariate that accurately represents the long-term trend by incorporating climate physics and anthropogenic influence through greenhouse gas emission scenarios. For this purpose, we consider the output from the Geophysical Fluid Dynamics Laboratory (GFDL) model GFDL-CM3 simulation \citep{shaltout2019recent} based on current century Coupled Model Intercomparison Project Phase 5 (CMIP5) scenarios RCP 4.5 and RCP 8.5. The GFDL-CM3 simulation outputs can be downloaded from the website \url{https://cds.climate.copernicus.eu}. While the GFDL-CM3 simulations are available over a $1^\circ \times 1^\circ$ grid (39 grid cells for the Red Sea), we here average the annual simulated SST for the entire Red Sea region, which yields only one value per year. Figure~\ref{fig:RCP} displays the SST projections based on RCP 4.5 and 8.5. 
\begin{figure}[t!]
\begin{center}
\adjincludegraphics[width = 0.65\linewidth, trim = {{.0\width} {.0\width} {.0\width} {.0\width}}, clip]{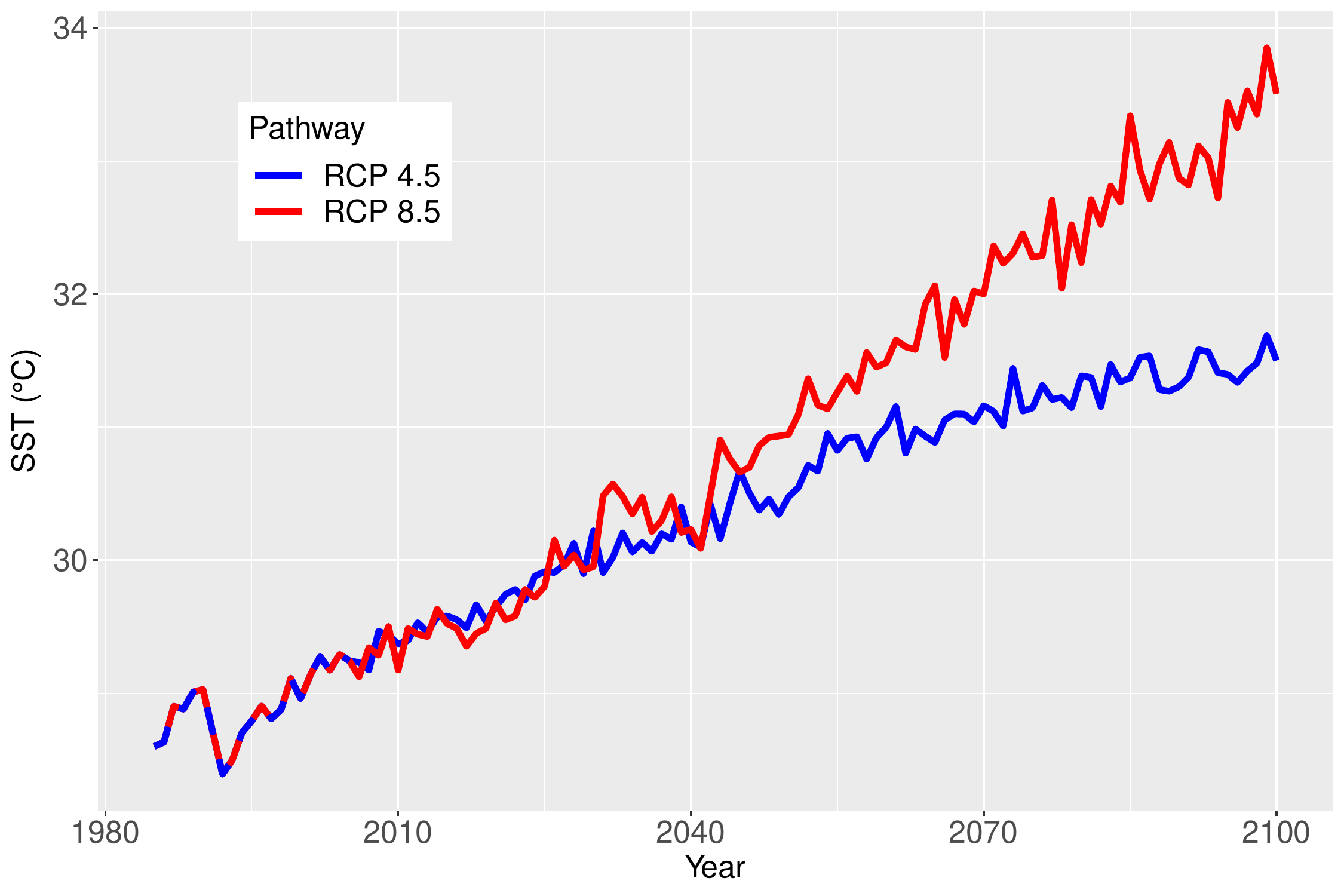}
\end{center}
\caption{Annual mean SST projections for the Red Sea, based on RCP 4.5 (blue) and RCP 8.5 (red) scenarios. The two RCPs are the same during the historic period (1985--2005) and different during the future period (2006-2100).}
\label{fig:RCP}
\end{figure}
We performed two separate analyses based on these two RCP scenarios. The two RCPs are the same during the historic period (1985--2005) and start to diverge during the future period (2006-2100). Here, we mainly focus on our findings based on the RCP 8.5 SST trajectory, which is sometimes referred to a ``business-as-usual'' scenario (without mitigation measures) and sometimes to a worst-case scenario. We thus consider it as a relatively pessimistic upper bound for SST projections, while RCP 4.5 is a more optimistic, moderate case scenario. For completeness, we also report the results for RCP 4.5 in the Supplementary Material.}

\begin{figure}[t!]
\begin{center}
	\adjincludegraphics[height = 0.4\linewidth, width = 0.31\linewidth, trim = {{.13\width} {.0\width} {.12\width} {.0\width}}, clip]{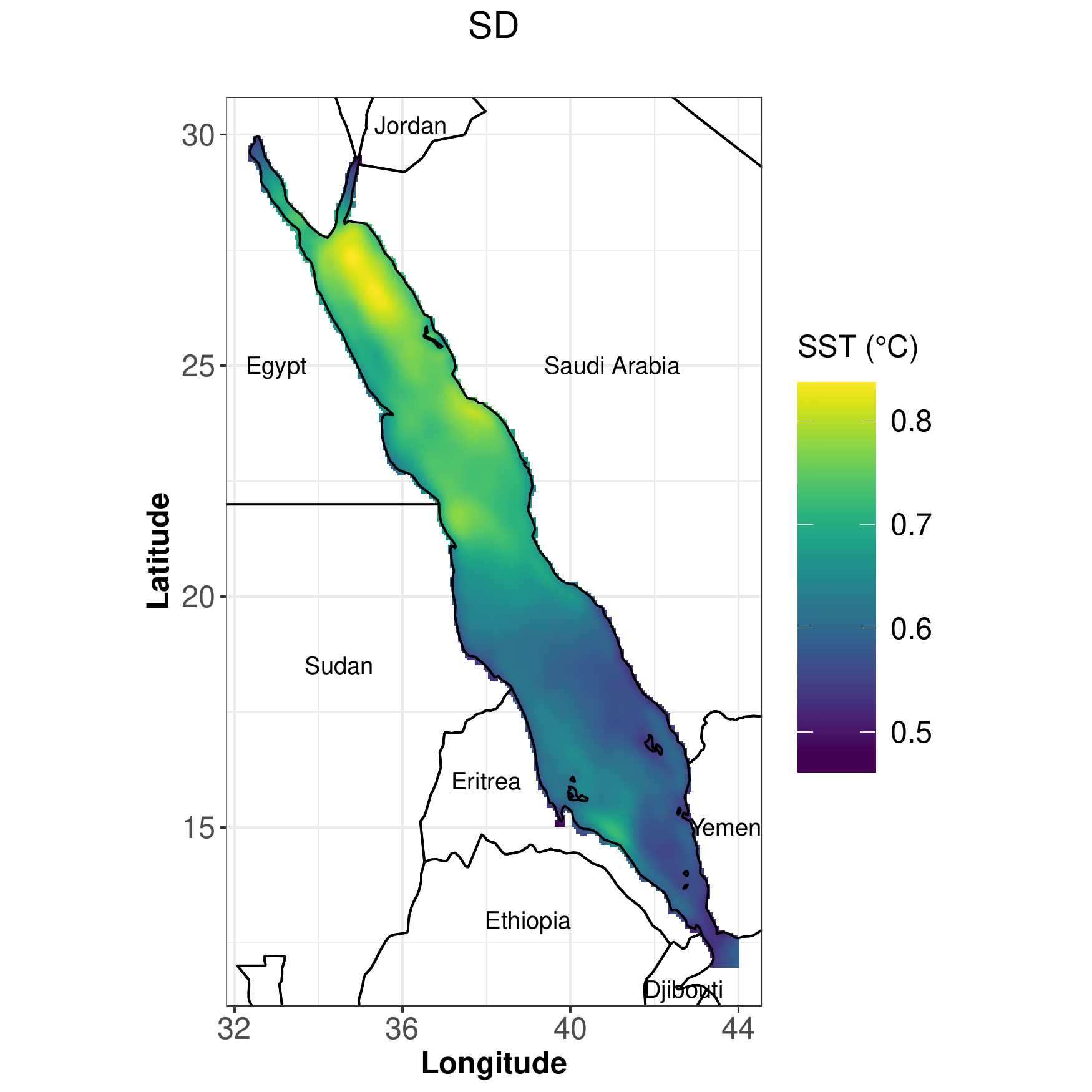}
	\adjincludegraphics[height = 0.4\linewidth, width = 0.31\linewidth, trim = {{.13\width} {.0\width} {.12\width} {.0\width}}, clip]{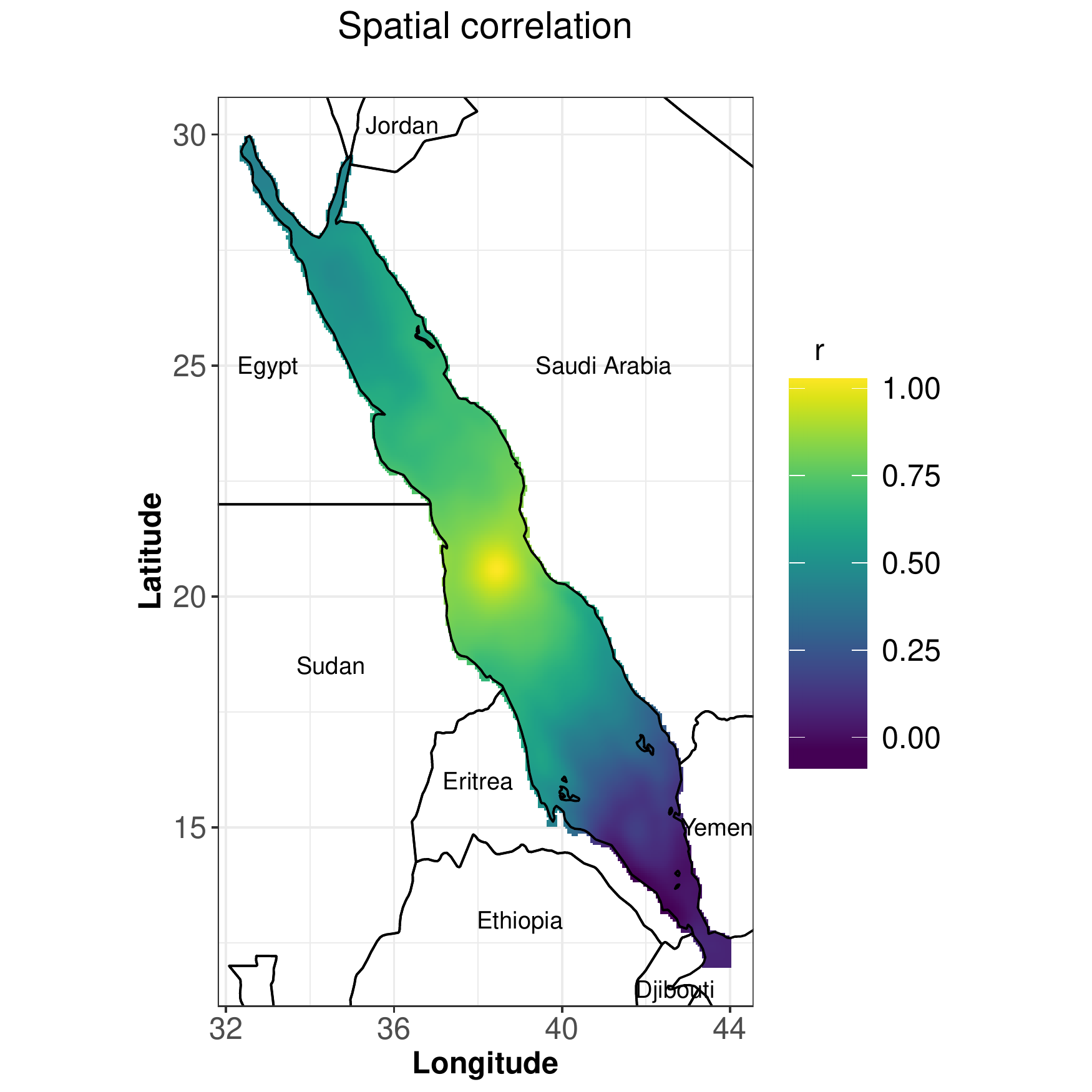}
	\adjincludegraphics[height = 0.4\linewidth, width = 0.31\linewidth, trim = {{.13\width} {.0\width} {.12\width} {.0\width}}, clip]{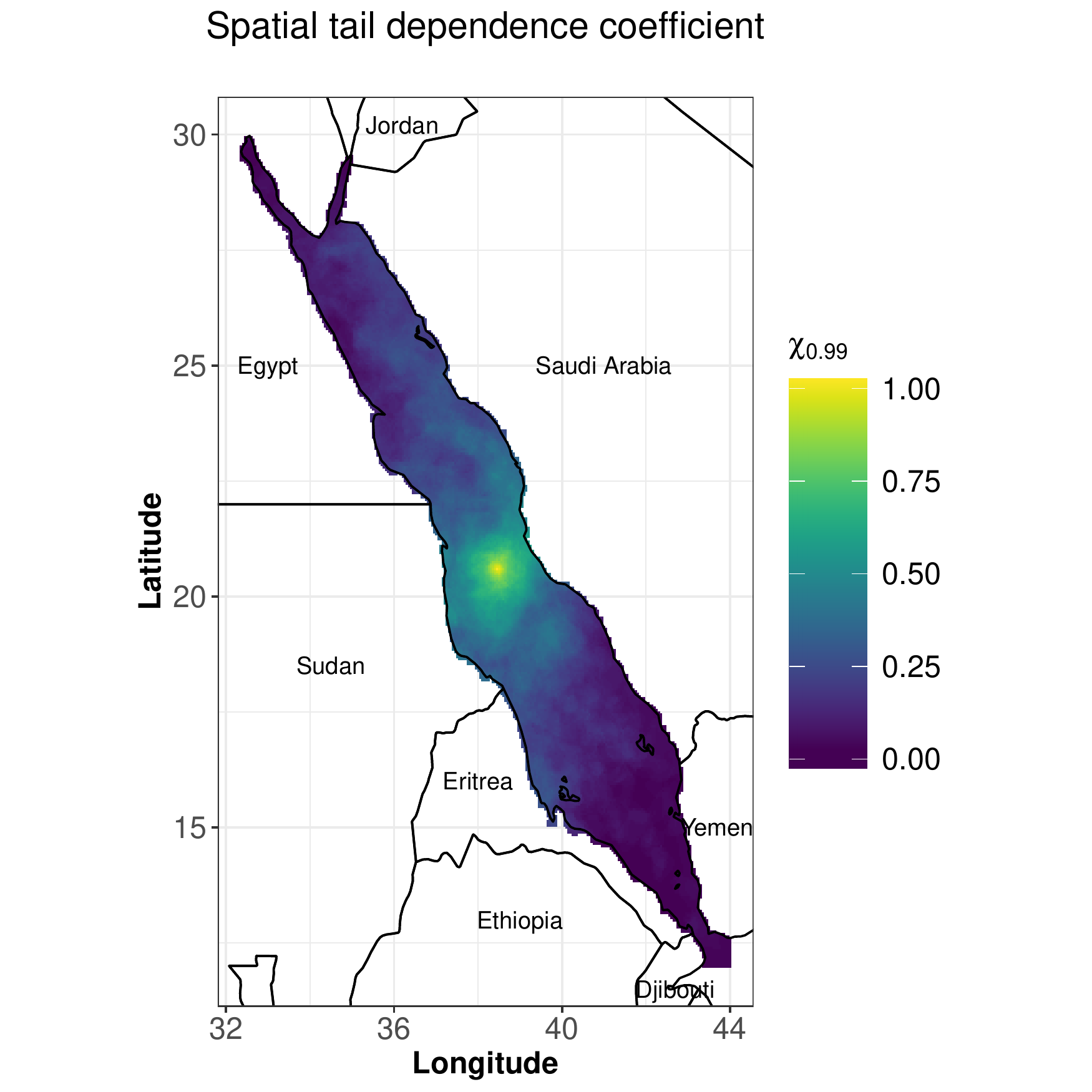}
\end{center}
\caption{Grid cell-wise SD (left), spatial correlation ($r$, middle) and extremal dependence ($\chi_{0.99}$, right) profiles corresponding to the grid cell ($38.48^\circ$E, $20.62^\circ$N), respectively. }
	\label{sd_corr_chi}
\end{figure}

We then compute the site-wise standard deviations (SDs) of the detrended SST data (with trend obtained by spline smoothing as discussed in \S \ref{mean_modeling}). 
The results are presented in the left panel of Figure \ref{sd_corr_chi}. Again, there is a highly nonstationary pattern, with high SDs near the northeast region and low SDs near the southeast region. The middle panel of Figure \ref{sd_corr_chi} shows the spatial correlation structure with respect to the central grid cell ($38.48^\circ$E, $20.62^\circ$N). The correlation values in the northern region are significantly higher than the values in the southern region despite being at the same distance. This suggests that the spatial correlation structure is also highly nonstationary.


To investigate the extremal dependence structure, { whose specification and estimation is crucial for hotspot detection,} we then compute the empirical tail dependence coefficient with respect to the same center grid cell ($38.48^\circ$E, $20.62^\circ$N). The tail dependence coefficient between two random variables $Y_1$ and $Y_2$ is defined as $\chi = \lim_{u \rightarrow 1} \chi_u$ {(provided the limit exists)}, where
\begin{eqnarray} \label{chi_def}
\chi_u =  \textrm{Pr} \left\{Y_1 > F_1^{-1}(u) \mid Y_2 > F_2^{-1}(u)\right\},
\end{eqnarray}
and $F_1$ and $F_2$ are the marginal distribution functions of $Y_1$ and $Y_2$, respectively. A nonzero value of $\chi$ indicates asymptotic dependence while $\chi = 0$ indicates asymptotic independence. Here, we estimate $\chi$ with the empirical conditional probability $\hat{\chi}_u$ with $u=0.99$. The values are reported on the right panel of Figure \ref{sd_corr_chi}. The observed values are nonzero throughout a major portion of the spatial domain indicating the necessity for a model that can capture nonzero extremal dependence (unlike GPs)  at large distances and high thresholds.

Finally, we investigate the bias in estimating high quantiles when fitting normal and Student's $t$ distributions. High biases observed at many grid cells indicate the need for a more flexible model than the usual parametric alternatives. As mixture models fit low through high quantiles of the distributions more flexibly, a semiparametric model is warranted.


More details on the exploratory analysis are provided in the Supplementary Material. To summarize, we need a model that accounts for spatially-varying trend and seasonality components within the mean structure, while the residual variability needs to be modeled through a mixture of spatial processes that allows for extremal dependence. Considering the high dimension, a low-rank approach is necessary to ensure that inference is practically feasible.



\section{Modeling}
\label{methodology}

\subsection{General framework}

Let $\mathcal{D}$ denote the spatial domain of the Red Sea. We model the Red Sea SST data as 
\begin{equation} \label{eq1}
    Y_t(\bm{s}) = \mu_t(\bm{s}) + \varepsilon_t(\bm{s}),
\end{equation}
where $Y_t(\bm{s})$ denotes the observed SST at  location $\bm{s} \in \mathcal{D}$ and at time $t \in \lbrace 1, 2, \ldots, T = 1612 \rbrace$, $\mu_t(\bm{s})$ is the mean SST profile and $\varepsilon_t(\bm{s})$ is the corresponding error component.

In \S\ref{mean_modeling}, we first discuss the modeling of the mean term $\mu_t(\bm{s})$, and in \S\ref{spatdep}, we then discuss the modeling of the residual process $\varepsilon_t(\bm{s})$ for $\bm{s} \in \mathcal{D}$ and $t \in \lbrace 1, 2, \ldots, T \rbrace$. In \S\ref{overall}, we specify the overall Bayesian model, and in \S\ref{properties}, we describe the model properties.

\subsection{Mean modeling}
\label{mean_modeling}
By an abuse of notation, we write $\mu_t(\bm{s}) = \mu(t_1, t_2, \bm{s})$ where $t_1 = \lceil t/52 \rceil$ denotes the year corresponding to time $t$ and $t_2 = t - 52 (t_1 - 1)$ denotes the corresponding week within the $t_1$-th year. Here, $t_1 \in \lbrace 1, \ldots, T_1 = 31 \rbrace$ for 31 years of data and $t_2 \in \lbrace 1, \ldots, T_2 = 52 \rbrace$ for the 52 weeks within each year. Henceforth, this one-to-one relation between $t$ and $(t_1, t_2)$ holds for the rest of the paper. In spite of the data being observed at discrete time points, we model the mean SST profile as a continuous function of $t_1$ and $t_2$. 



{ Let $x^*_{t_1}$ be the mean simulated SST based on a RCP scenario for the $t_1$-th year. We assume that $\mu(t_1, t_2, \bm{s})$ is linear in $x^*_{t_1}$, and we define $\mu(t_1, t_2, \bm{s}) = \beta_1(t_2, \bm{s}) x_{t_1, 1}^{(0)} + \beta_2(t_2, \bm{s}) x_{t_1, 2}^{(0)}$, where $x_{t_1, 2}^{(0)}$ is a standardized version of $x^*_{t_1}$, and $\beta_1(t_2, \bm{s}),\beta_2(t_2, \bm{s})$ denote the regression coefficients (intercept and slope with respect to $x^*_{t_1}$, respectively) that vary over space, as well as for each week of a specific year. Furthermore, we write the regression coefficients $\beta_{p_0}(t_2, \bm{s}),p_0=1,2$, as $\beta_{p_0}(t_2, \bm{s}) = \sum_{p_1=1}^{P_{\mathcal{T}}} \beta_{p_0, p_1}(\bm{s}) x_{t_2, p_1}^{(1)}$, where $x_{t_2, p_1}^{(1)},p_1 = 1, \ldots, P_{\mathcal{T}}$, are cubic B-splines defined over the continuous interval $[1, T_2]$ with equidistant knots and evaluated at $t_2$ and thus, the seasonal variation is modeled as a smooth function of week within each year. Considering one B-spline per month, we have $P_{\mathcal{T}} = 12$ that helps capturing the monthly-varying features. Finally, let the grid cells within $\mathcal{D}$ be denoted by $\bm{s}_1, \ldots, \bm{s}_N$. In order to reduce the computational burden due to the high spatial dimension $N=16703$, we consider a low-rank approximation of the spatially-varying coefficients by specifying $\beta_{p_0, p_1}(\bm{s}_n) = \sum_{p_2=1}^{P_{\mathcal{S}}} \beta_{p_0,p_1,p_2} x^{(2)}_{n, p_2}$, where $x^{(2)}_{n, p_2}$ are suitable spatial basis functions. While other choices are also possible, we consider tensor products of cubic B-splines defined over a rectangular surface covering $\mathcal{D}$. A large number of basis functions $P_{\mathcal{S}}$ provides a lot of local detail, but it also increases the computational burden. On the contrary, a small $P_{\mathcal{S}}$ is computationally appealing but might over-smooth the spatial mean surfaces. In practice, this trade-off depends on the computational resources available and the required modeling accuracy.

To be more precise, we now discuss further technical details regarding the design matrices. Let $\bm{X}_0$ denote the $(T_1 \times 2)$-dimensional matrix with $(t_1, p_0)$-th entry denoted by $x^{(0)}_{t_1, p_0}$.
We standardize covariates such that $x_{t_1, 1}^{(0)} =  T_1^{-1/2}$ and $x_{t_1, 2}^{(0)} = {(x^*_{t_1} - \bar{x}^*)}/{\{\sum_{t_1^* = 1}^{T_1} (x^*_{t_1} - \bar{x}^*)^2\}^{1/2}}$, with $\bar{x}^* = T_1^{-1}\sum_{t_1^* = 1}^{T_1} x^*_{t_1}$. This ensures that $\bm{X}_0$ is an orthonormal matrix which is important from a computational perspective. For the low-rank structure of $\beta_{p_0}(t_2, \bm{s})$, we consider the same B-spline basis functions for $p_0 = 1,2$ for convenience, though other choices are also possible. We denote the $(T_2 \times P_{\mathcal{T}})$-dimensional design matrix corresponding to the seasonal effects by $\bm{X}_1$, with $(t_2, p_1)$-th entry $x_{t_2, p_1}^{(1)}$. For choosing suitable basis functions for the $\beta_{p_0, p_1}(\bm{s}_n)$'s, taking the elongated geometry of the Red Sea into account, we place 30 equidistant B-splines along the northwest--southeast direction and 10 equidistant B-splines along the southwest--northeast direction. Out of these 300 B-splines, we only keep the $P_{\mathcal{S}} = 189$ of them that represent more than 99\% of the total weight over $\bm{s}_1, \ldots, \bm{s}_N$. The knots are well spread across all the Red Sea and hence the splines are able to capture local characteristics quite well. We denote the corresponding $(N \times P_{\mathcal{S}})$-dimensional design matrix by $\bm{X}_2$, with $(n, p_2)$-th entry $x^{(2)}_{n, p_2}$.}

In summary, we model the mean SST at spatial location $\bm{s}_n$ and time point $t$ as
\begin{eqnarray} \label{mean_full}
 \mu_{t}(\mathbf{s}_n) &=& \sum_{p_0=1}^{2} \sum_{p_1=1}^{P_{\mathcal{T}}} \sum_{p_2=1}^{P_{\mathcal{S}}}  \beta_{p_0,p_1,p_2} x^{(0)}_{t_1, p_0} x^{(1)}_{t_2, p_1} x^{(2)}_{n, p_2}.
 \end{eqnarray}

Let the spatial mean vector at time $t$ be $\bm{\mu}_t = \left[\mu_t(\bm{s}_1), \ldots, \mu_t(\bm{s}_N)\right]'$, and combining all time points, let $\bm{\mu} = [\bm{\mu}'_1, \ldots, \bm{\mu}_T']'$. Grouping the regression coefficients, let $\bm{\beta}_{p_0, p_1} = \left[\beta_{p_0,p_1,1}, \ldots, \beta_{p_0,p_1,P_{\mathcal{S}}} \right]'$ and $\bm{\beta}_{p_0} = [\bm{\beta}_{p_0, 1}', \ldots, \bm{\beta}_{p_0, P_{\mathcal{T}}}']'$. Denoting the two columns of $\bm{X}_0$ by $\bm{X}_{0;1}$ and $\bm{X}_{0;2}$, respectively, we can write $\bm{\mu}$ in vectorial form as $\bm{\mu} = \left[\bm{X}_{0;1} \otimes \bm{X}_1 \otimes \bm{X}_2\right] \bm{\beta}_1 + \left[\bm{X}_{0;2} \otimes \bm{X}_1 \otimes \bm{X}_2\right] \bm{\beta}_2$, where $\otimes$ denotes the Kronecker product between two matrices.

\subsection{Spatial dependence modeling}
\label{spatdep}
We now discuss the modeling of the residual process $\varepsilon_t(\bm{s})$. We here assume that the $\varepsilon_t(\bm{s})$'s are iid across time, with zero mean, and we write $\varepsilon(\bm{s})$ for a generic copy of $\varepsilon_t(\bm{s})$. Our semiparametric Bayesian model for the spatial residual processes $\varepsilon_t(\cdot)$ is based on a Dirichlet process mixture (DPM) of parametric low-rank Student-$t$ processes (LTPs). We first describe the construction of LTPs, and then discuss the modeling based on mixtures.

\subsubsection{Low-rank Student-$t$ process (LTP)}
LTPs are richer than LGPs as they have heavier marginal and joint tails, and they can capture spatial extremal dependence contrary to Gaussian processes. At a spatial location $\bm{s}$, we model a realization from an LTP as

\begin{eqnarray} \label{lmt_def}
 && \varepsilon(\bm{s}) = \sigma \left\{\bm{h}'(\bm{s}) \bm{Z} + \eta(\bm{s}) \right\},\quad \bm{s}\in\mathcal D,
\end{eqnarray}
where $\bm{h}(\bm{s})$ denotes the vector of length $L$ comprised of some spatial basis functions evaluated at $\bm{s}$. The random effects are specified as $\bm{Z} \sim \textrm{Normal}_L\left( \bm{0}, \bm{\Phi}\right)$ for some positive definite matrix $\bm{\Phi}$, while $\sigma^2 \sim \textrm{Inverse-Gamma}\left( \frac{a}{2}, \frac{a}{2} - 1 \right)$, $a>2$, and $\eta(\cdot)$ denotes a spatial white noise process (i.e., nugget effect) such that $\eta(\bm{s}) \overset{iid}{\sim} \textrm{Normal}(0, \tau^2)$, $\tau>0$.

For the $N$ spatial grid cells $\bm{s}_1, \ldots, \bm{s}_N$, let $\bm{\varepsilon} = \left[\varepsilon(\bm{s}_1), \ldots, \varepsilon(\bm{s}_N) \right]'$ be the vector of observed values from the process $\varepsilon(\cdot)$. Moreover, let $\bm{H}$ be the $(N \times L)$-dimensional matrix, whose columns are the different spatial basis functions evaluated at $\bm{s}_1, \ldots, \bm{s}_N$. After marginalization over the random effects $\bm{Z}$ and $\sigma^2$, the joint distribution of $\bm{\varepsilon}$ is 
\begin{eqnarray} \label{lmt}
\bm{\varepsilon} &\sim& T_a\left( \bm{0}_N, \frac{a - 2}{a} \left(\bm{H} \bm{\Phi} \bm{H}' + \tau^2 \bm{I}_N \right)  \right),
\end{eqnarray}
where $T_{\tilde{a}}(\tilde{\bm\mu}, \tilde{\bm{\Sigma}})$ denotes the multivariate Student's $t$ distribution with location vector $\tilde{\bm\mu}$, dispersion matrix $\tilde{\bm{\Sigma}}$ and degrees of freedom $\tilde{a}$, $\bm{I}_N$ is the $N$-by-$N$ identity matrix, and $\bm{0}_N$ is the zero vector of length $N$. In case (temporal) replications are available from the spatial process $\varepsilon(\cdot)$, { a rough estimate of} the spatial covariance matrix of $\bm{\varepsilon}$ (which exists since $a>2$) can be { obtained upfront} using the sample covariance matrix $\hat{\bm{\Sigma}}=\hat{\textrm{Cov}}(\bm{\varepsilon})$, {with $\varepsilon$ obtained from a preliminary least-squares fit of model \eqref{eq1} based on the specification \eqref{mean_full} after subtracting the fitted mean from the data}. Here, we consider spatial basis functions to be the $L\ll N$ eigenvectors of $\hat{\bm{\Sigma}}$ with the largest corresponding eigenvalues. In other words, the matrix $\bm{H}$ is comprised of empirical orthogonal functions (EOFs). Specifically, let $\bm{\Delta}$ be the diagonal matrix with the $L$ largest eigenvalues of $\hat{\bm{\Sigma}}$, and $\bm{H}$ be the matrix with column vectors equal to the corresponding eigenvectors. Then, we have the low-rank approximation $\hat{\bm{\Sigma}} \approx \bm{H} \bm{\Delta} \bm{H}'$, {where the right-hand side has rank $L$}. Other choices of basis functions $\bm{h}(\bm{s})$ are also possible (even in case replicates are unavailable); a detailed discussion about the choice of $\bm{h}(\bm{s})$ is provided in \cite{wikle2010low}. { For illustration, the six most important EOFs for the Red Sea data are displayed in the Supplementary Material. They reveal interesting geographic patterns, which indicates that our EOF-based approach makes sense from a practical perspective. While obtaining the $(N \times N)$-dimensional matrix $\hat{\bm{\Sigma}}$ is computationally intensive, we here need to compute it only once upfront (before running the MCMC algorithm) and the function \code{cova} from the R package \code{Rfast} \citep{papadakis2017rfast} can be exploited to speed up the computation. Furthermore, obtaining all the eigenvalues and the corresponding eigenvectors would also be time-consuming, but the function \code{eigs\_sym} from the R package \code{rARPACK} \citep{qiu2016package} can be used to very efficiently compute a small number of highest eigenvalues and the corresponding eigenvectors.} While $\bm{\Phi} = \bm{\Delta}$ could be assumed { to be fixed}, we consider instead $\bm{\Phi}$ in \eqref{lmt} to be unknown and use an informative prior on $\bm{\Phi}$ with prior mean equal to $\bm{\Delta}$.  The nugget component $\bm{\eta}=\left[\eta(\bm{s}_1), \ldots, \eta(\bm{s}_N) \right]'$ is important for $\textrm{Cov}(\bm{\varepsilon})$ to be full-rank, though $\tau^2$ is expected to be small. The specific parametrization of $\sigma^2$ ensures that $\textrm{Cov}(\bm{\varepsilon}) = \bm{H} \bm{\Phi} \bm{H}' + \tau^2 \bm{I}_N$, so plugging the prior mean of $\bm{\Phi}$, we get the approximation $\textrm{Cov}(\bm{\varepsilon}) \approx \bm{H} \bm{\Delta} \bm{H}' + \tau^2 \bm{I}_N \approx \hat{\bm{\Sigma}}$. Hence, {overall}, we construct a zero-mean LTP, where the covariance structure resembles the sample covariance { estimated in a preliminary step}.

\subsubsection{Dirichlet process mixture (DPM) of LTPs}
In case we do not have any (temporal) replicates of the process $\varepsilon(\cdot)$, parametric assumptions are required (though $\hat{\bm{\Sigma}}$---and hence $\bm{H}$---are not available and another choice of basis functions is required). However, for the Red Sea data, independent temporal replicates of the anomalies, $\varepsilon_t(\bm{s})$, $t=1,\ldots,T$, are available and we can thus estimate the underlying spatial process semiparametrically.

Considering that our focus mainly lies in inferences from the tail, we here extend the construction (\ref{lmt_def}) by modeling the residuals using a DPM in the same spirit as \citet{hazra2018semiparametric}, where the characteristics of the bulk and the tail of the anomaly process are described by different mixture components with component-specific parameters. Thus, our approach can automatically and probabilistically cluster observations into weeks characterized by ``normal conditions'' or weeks characterized by ``abnormal (extreme) conditions'' , without any artificial and subjective thresholding. Thus, tail inference is expected to be minimally influenced by observations from the bulk, while providing a reasonable fit on the entire probability range.
 
Now, we assume that for all $t$, $\bm{\varepsilon}_t = \left[\varepsilon_t(\bm{s}_1), \ldots, \varepsilon_t(\bm{s}_N) \right]'$ are \textit{iid} $N$-dimensional realizations from a LTP-DPM model with $K$ mixture components for some natural number $K$. The corresponding multivariate density function is
	\begin{equation} \label{dpdensity}
     f_{\textrm{DPM}}(\bm{\varepsilon}) = \sum_{k=1}^{K} \pi_k f_{T}(\bm{\varepsilon}; \bm{\Theta}_k),
	\end{equation}
where $\pi_k > 0$ are the mixture probabilities with $\sum_{k=1}^{K} \pi_k = 1$, $\bm{\Theta}_k$ denotes the set of parameters of the $k$-th LTP component, and $f_{T}(\cdot)$ denotes the density of an $N$-dimensional realization from the LTP in (\ref{lmt}). 
When $K = \infty$, the model becomes fully nonparametric. 

The main advantage of the LTP-DPM model lies in its hierarchical Bayesian model representation. The model can be rewritten as a clustering model, where, conditional on the random cluster label $g_t$ with probability mass function $\textrm{Pr}(g_t = k) = \pi_k$, $k \in \lbrace 1, \ldots, K \rbrace$, we have $\bm{\varepsilon}_t \sim f_{T}(\cdot\mid \bm{\Theta}_{g_t})$. Thus, our LTP-DPM model assumes that weeks with similar residuals can be clustered together, their distribution being described by the same LTP. Treating the cluster labels $g_t$ as unknown, the model accounts for uncertainty in cluster allocation.

We assign a truncated stick-breaking prior distribution \citep{sethuraman1994constructive} on the mixture probabilities, $\pi_k$, $k\in\{1, \ldots, K\}$. 
Precisely, we have the following constructive representation: $\pi_1 = V_1 \sim \textrm{Beta}(1, \delta)$ for some Dirichlet process concentration parameter $\delta > 0$, and subsequently, $\pi_k = (1-\sum_{i=1}^{k-1}\pi_i)V_k$ for $k=1, \ldots,K-1$ with $V_k \overset{\textrm{iid}}{\sim} \textrm{Beta}(1, \delta)$. As we consider $K$ to be finite, we set $V_K = 1$ so that $\sum_{k=1}^{K} \pi_k = 1$. In our MCMC implementation, we exploit the one-to-one correspondence between the $\pi_k$'s and the $V_k$'s, by iteratively updating the latter to estimate the former. We write $\bm{\pi} = [\pi_1, \ldots, \pi_K]' \sim \textrm{Stick-Breaking}(\delta)$.


\subsection{Overall model}
\label{overall}
Considering two types of spatial basis functions within $\mu_t(\bm{s})$ and $\varepsilon_t(\bm{s})$ leads to collinearity issues between fixed and random effects. 
To resolve this issue, we divide the matrix $\bm{X}_2$ from \S\ref{mean_modeling} into two parts based on its projection onto the column space the matrix $\bm{H}$ from \S\ref{spatdep}. The projection matrix is $\bm{P}_{\bm{H}} = \bm{H}\left(\bm{H}'\bm{H}\right)^{-1}\bm{H}' = \bm{H}\bm{H}'$, as $\bm{H}$ is an orthonormal matrix. We have $\bm{X}_2 = \bm{X}_{2;1} + \bm{X}_{2;2}$ where $\bm{X}_{2;1} = \bm{P}_{\bm{H}} \bm{X}_2$, and $\bm{X}_{2;2} = (\bm{I}_N - \bm{P}_{\bm{H}}) \bm{X}_2$, and then rewrite $\bm{\mu} = \sum_{i=1}^{2} \sum_{j=1}^{2} \left(\bm{X}_{0;i} \otimes \bm{X}_1 \otimes \bm{X}_{2;j}\right) \bm{\beta}_{i;j}$ with two separate vectors of coefficients corresponding to $\bm{X}_{2;1}$ and $\bm{X}_{2;2}$ for each of the intercept-related and slope-related terms. For a specific grid cell $\bm{s}_n$ and time point $t$, $\mu_t(\bm{s}_n) = \sum_{i=1}^{2} \sum_{j=1}^{2} x^{(0;i)}_{t_1}  \left(\bm{x}^{(1)}_{t_2} \otimes \bm{x}_n^{(2;j)}\right) \bm{\beta}_{i;j}$, where $x^{(0;i)}_{t_1}$ is the $t_1$-th entry of $\bm{X}_{0;i}$, $\bm{x}^{(1)}_{t_2}$ is the $t_2$-th row of $\bm{X}_1$, and $\bm{x}_n^{(2;j)}$ is the $n$-th row of $\bm{X}_{2;j}$.

Overall, our hierarchical model is defined as follows. Given the cluster label $g_t = k$,
\begin{eqnarray} \label{full_model}
\nonumber && Y_t(\bm{s}) = \mu_t(\bm{s}) +  \sigma_t \left\{\bm{h}'(\bm{s}) \bm{Z}_t + \eta_t(\bm{s}) \right\}, \\
\nonumber && \bm{Z}_t \sim \textrm{Normal}_L\left(\bm{0}, \bm{\Phi}_k \right),\quad \eta_t(\bm{s}) \overset{iid}{\sim} \textrm{Normal}\left(0, \tau_k^2 \right), \\
 && \sigma_t^2 \sim \textrm{Inverse-Gamma}\left( \frac{a_k}{2}, \frac{a_k}{2} - 1 \right),
\end{eqnarray}
where $\mu_t(\bm{s})$ is the mean SST profile, $\bm{h}(\bm{s})$ is the vector of spatial basis functions as described in \S \ref{spatdep}, and $\bm{Z}_t$ and $\sigma_t$ denote independent copies of the corresponding random effects. The set of parameters of the LTP  corresponding to time $t$ is $\bm{\Theta}_{k} = \lbrace \bm{\Phi}_k, \tau_k^2, a_{k} \rbrace$. We treat the cluster-specific parameters $\bm{\Theta}_{k}$ as unknown and put hyperpriors on them. We assume that $\bm{\Theta}_{k} \overset{\textrm{iid}}{\sim} G_{\bm{\Theta}}$, where the components of $\bm{\Theta}_{k}$, i.e., $\bm{\Phi}_k, \tau_k^2$ and $a_{k}$, are treated as independent of each other. Our choice of hyperpriors is discussed in \S\ref{prior_choices}.

\subsection{Model properties}
\label{properties}

For model (\ref{full_model}), the conditional mean and covariance structure of $Y_t(\bm{s}_n)$ given the coefficients $\bm{\beta}_{i;j}$, and the cluster-specific parameters $\bm{\Theta}_k$ are
\begin{eqnarray} \label{meancovprop}
\nonumber && \textrm{E}\left\{Y_t(\bm{s}_n) \mid \bm{\beta}_{i;j}, i,j=1,2 \right\} = \mu_t(\bm{s}_n) = \sum_{i=1}^{2} \sum_{j=1}^{2} x^{(0;i)}_{t_1}  \left(\bm{x}^{(1)}_{t_2} \otimes \bm{x}_n^{(2;j)}\right) \bm{\beta}_{i;j}, \\
 && \textrm{Cov}\left\{Y_t(\bm{s}_{n_1}), Y_t(\bm{s}_{n_2}) \mid \bm{\Theta}_k; k =1, \ldots,K \right\} = \sum_{k=1}^{K} \pi_k \left(\bm{h}_{n_1} \bm{\Phi}_k \bm{h}'_{n_2} + \tau_k^2 \mathbb{I}_{\lbrace n_1 = n_2 \rbrace} \right),
\end{eqnarray}
where $\bm{h}_n$ denotes the $n$-th row of $\bm{H}$ and $\mathbb{I}_{\lbrace n_1 = n_2 \rbrace} = 1$ if $n_1 = n_2$ and zero otherwise.

The mean structure is clearly nonstationary both in space and time, non-linear and includes interaction terms between the spatial and temporal effects. Thus, the model can capture spatially-varying seasonality and trend as well as seasonally-varying trends at each spatial grid cell. Because we specify high-resolution B-spline knot locations both over space and time, our model can capture the local features in the mean behavior reasonably well. 

The covariance between the observations at grid cells $\bm{s}_{n_1}$ and $\bm{s}_{n_2}$ is dependent on both $\bm{s}_{n_1}$ and $\bm{s}_{n_2}$ and cannot be reduced to a function of the spatial lag $\bm{s}_{n_1} - \bm{s}_{n_2}$. Hence, the covariance structure is also nonstationary. For the Dirichlet process atoms $\bm{\Phi}_k$, $k=1, \ldots, K$, we consider priors so that $\textrm{E}(\bm{\Phi}_k) = \bm{\Delta}$; recall \S\ref{spatdep}. Marginalizing over $\bm{\Phi}_k$, we get $\textrm{Cov}\left\{Y_t(\bm{s}_{n_1}), Y_t(\bm{s}_{n_2}) \right\} = \bm{h}_{n_1} \bm{\Delta} \bm{h}'_{n_2} + \tau^2 \mathbb{I}_{\lbrace n_1 = n_2 \rbrace}$, where $\tau^2 = \sum_{k=1}^{K} \pi_k \tau_k^2$. Considering $\tau^2$ to be small, we get the approximation $\textrm{Cov}(\bm{Y}_t) \approx \hat{\bm{\Sigma}}$, where $\bm{Y}_t = [Y_t(\bm{s}_1), \ldots, Y_t(\bm{s}_N)]'$ and $\hat{\bm{\Sigma}}$ is as discussed in \S\ref{spatdep}. Thus, the LTP-DPM model is centered around a low-rank Student's $t$ process, constructed as in \S\ref{spatdep} with mean structure discussed in \S\ref{mean_modeling}.


Considering $\varepsilon_t(\cdot)$ as an infinite-dimensional spatial process and ignoring the nugget term, for $L = \infty$, we can write $\varepsilon_t(\bm{s}) = \sum_{l=1}^{\infty} Z_{lt} h_l(\bm{s})$. For any pair $l_1$ and $l_2$ with $l_1 \neq l_2$, the  vector $[Z_{l_1 t}, Z_{l_2 t}]'$ follows a DPM of bivariate zero-mean Student's $t$ distributions which spans any bivariate zero-mean distribution when $K=\infty$. {Integrating with respect to the priors of $\bm{\Phi}_k$, $\textrm{Cov}(Z_{l_1 t}, Z_{l_2 t}) = 0$. Thus, for $L,K = \infty$, the covariance of the proposed model satisfies the criteria of the Karhunen-Lo\`eve Theorem \citep{alexanderian2015brief} and spans the covariance structure of all squared-integrable stochastic processes with continuous covariance functions. Under suitable regularity conditions, 
posterior consistency of the proposed model holds \citep{wu2010l1, ghosal2017fundamentals}.}


The spatial tail dependence coefficient between two different grid cells $\bm{s}_{n_1}$ and $\bm{s}_{n_2}$, defined in (\ref{chi_def}), is nonstationary for the proposed LTP-DPM model and it is easy to show that it is determined by the heaviest-tailed mixture component in \eqref{dpdensity}. Using the tail properties of Student-$t$ copulas \citep{Demarta.McNeil:2005}, its expression may be written explicitly as 
$\chi(\bm{s}_{n_1}, \bm{s}_{n_2}) = 2 \bar{F}_T\left(\sqrt{(a_m + 1) \{1 - r_m(\bm{s}_{n_1}, \bm{s}_{n_2})\}/\{1 + r_m(\bm{s}_{n_1}, \bm{s}_{n_2})\}}; 0, 1, a_m + 1 \right)$,
where $m = \arg \min_k \left \lbrace a_k\right \rbrace$, and $r_m(\bm{s}_{n_1}, \bm{s}_{n_2}) = \bm{h}_{n_1} \bm{\Phi}_m \bm{h}'_{n_2} / \sqrt{ \prod_{\tilde{n}=n_1,n_2} (\bm{h}_{\tilde{n}} \bm{\Phi}_m \bm{h}'_{\tilde{n}} + \tau_m^2)}$ denotes the underlying spatial correlation of the Gaussian term characterizing the $m$-th mixture component, and $\bar{F}_T(\cdot~; 0,1,a) = 1 - F_T(\cdot~; 0,1,a)$ is the survival function for a standard (univariate) Student's $t$ distribution with $a$ degrees of freedom; see, also \cite{hazra2018semiparametric}. Here, $\chi(\bm{s}_{n_1}, \bm{s}_{n_2})>0$ for any pair of sites, so our model can capture asymptotic dependence unlike Gaussian processes, and is an increasing function of $r_m(\bm{s}_{n_1}, \bm{s}_{n_2})$. {Asymptotic dependence holds, however, even when $r_m(\bm{s}_{n_1}, \bm{s}_{n_2}) = 0$ (unless $a_m \to \infty$). Therefore, the model cannot capture full independence, and dependence persists at large distances, even in the tails.} This is a downside of the proposed model in case the spatial domain is large and the process of interest is rough across the domain (e.g., with precipitation or wind speed data), but this should not be a big limitation for our Red Sea SST data, which remain strongly dependent at large distances. { From Equations \eqref{meancovprop} and the expression for $\chi(\bm{s}_{n_1}, \bm{s}_{n_2})$ above, we can see that the dependence structure (both covariance, and tail dependence) is nonstationary and is not impacted by the smooth spatiotemporal mean structure.}


\section{Bayesian inference and identification of hotspots}
\label{computation}



\subsection{Hyperpriors and an efficient Gibbs sampler}
\label{prior_choices}

We draw posterior inference about the model parameters in our model using Markov chain Monte Carlo (MCMC) sampling. Except for the degrees of freedom parameters, $a_k$, $k=1,\ldots,K$, of the LTP components of the LTP-DPM model, conjugate priors exist for all other model parameters, which allows Gibbs sampling. While Metropolis--Hastings sampling would be possible for the $a_k$'s, we prefer to consider discrete uniform priors on a fine grid of values, which allows drawing samples from the full conditional distributions in a fast and easy manner. This strategy is often considered in the literature due to its numerical stability; for example, \cite{gelfand2005bayesian} use it for posterior sampling from the range parameter of the spatial Mat\'ern covariance; see also \cite{morris2017space} 
and \cite{hazra2019multivariate}. 

For the vectors of fixed effects involved within the mean terms, $\bm{\beta}_{i;j}$, $i=1,2, j=1,2$, we consider the priors $\bm{\beta}_{i;j} \sim \textrm{Normal}_P( \mu_{i;j} \bm{1}_P, \sigma^2_{i;j} \bm{I}_P)$, where $\bm{1}_P$ is the $P$-dimensional vector of ones and $\bm{I}_P$ is the $P$-by-$P$ identity matrix, with $P = P_{\mathcal{T}}  P_{\mathcal{S}} = 2268$. {  While the posterior distribution of $\bm{\beta}_{i;j}$ is also a $P$-variate normal distribution, we exploit the Kronecker product structure of the high-dimensional posterior covariance matrix to avoid its computationally challenging inversion.} For the hyper-parameters $\mu_{i;j}$, we consider the relatively  non-informative priors $\mu_{i;j} \sim \textrm{Normal}(0, s^2_{i;j})$ with $s_{1;j} = 10^2$ and $s_{2;j} = 10$ for $j=1,2$. The parameter vectors $\bm{\beta}_{1;j}$, $j=1,2$, correspond to the intercept term, while $\bm{\beta}_{2;j}$, $j=1,2$, correspond to the slopes with respect to the simulated RCP based SST estimates. The absolute values of the intercept-related terms are likely to be large, while the slope terms are likely to be small and thus, we consider flatter prior for the $\mu_{1;j}$'s. For the hyper-parameters $\sigma_{i;j}^2$, we consider the priors $\sigma_{i;j}^2 \sim \textrm{Inverse-Gamma}(a_{i;j}, b_{i;j})$, $i=1,2,j=1,2$. We fix the hyper-parameters to $a_{1;j}=b_{1;j} = 0.01$ and $a_{2;j}=b_{2;j} = 0.1$ for $j=1,2$. While both priors are quite non-informative, we choose the hyper-priors differently following a similar logic as the one used when considering the priors for the $\mu_{i;j}$'s.


The parameters involved in the distribution of the error terms $\varepsilon_t(\bm{s}_n)$ are the component-specific parameters and hyper-parameters of the DPM model described in \S\ref{spatdep}. For the purpose of computation, we fix the number of components in the stick-breaking prior by setting $V_K = 1$ for some finite integer $K$. The choice of $K$ is problem-specific, and leads to a bias--variance trade-off. Large $K$ is desirable to increase the model flexibility (i.e., decrease the bias), but considering $K$ to be very large may lead to spurious estimates as the sampling from the parameters of a LTP component depend on the observations from that specific cluster and a large $K$ may lead to very few observations within some clusters (thus increasing the variance). In our application, we fit different models with $K=1,5,10$ and compare them by cross-validation (see \S\ref{crossvalidation}). The prior choices for the DPM model parameters are $\bm{\Phi}_k \overset{\textrm{iid}}{\sim} \textrm{Inverse-Wishart}\left(L+2, \bm{\Delta}\right)$ where $\bm{\Delta}$ is the diagonal containing the $L$ largest eigenvalues of $\hat{\bm{\Sigma}}$ (recall \S\ref{spatdep}), $\tau^2_k  \overset{\textrm{iid}}{\sim} \textrm{Inverse-Gamma}(1, 1)$, and $a_k \overset{\textrm{iid}}{\sim} \textrm{Discrete-Uniform}(2.1, 2.2, \ldots, 40.0)$. The prior for $\bm{\Phi}_k$ is conjugate and it ensures that $\textrm{E}(\bm{\Phi}_k) = \bm{\Delta}$, {although the variance of its elements is infinite}. We choose hyper-parameters for the prior of $\tau^2_k$ so that the mass is mainly distributed near zero. The existence of a continuous conjugate prior for $a_k$ is unknown and the discrete uniform prior avoids the need of Metropolis--Hastings sampling, as mentioned above. 
The support of the distribution considered here covers a wide range of degrees of freedom. The smallest values within the support of $a_k$ correspond to a heavy-tailed and strongly dependent process, while the largest values practically correspond to a near-Gaussian behavior. For the mixing probabilities $\pi_k$, we consider a stick-breaking prior as discussed in \S\ref{spatdep}. The distribution of $\bm{\pi}= [\pi_1, \ldots, \pi_K]$ involves the unknown concentration hyper-parameter $\delta$. We here consider a fairly non-informative conjugate  hyper-prior, $\delta \sim \textrm{Gamma}(0.1, 0.1)$. 

The full conditional distributions required for model fitting and prediction are provided in the Supplementary Material. In our data application, we implement the MCMC algorithm in \texttt{R} (\url{http://www.r-project.org}). We generate 60,000 posterior samples and discard the first 10,000 iterations as burn-in period. Subsequently, we thin the Markov chains by keeping one out of five consecutive samples and thus, we finally obtain $B=10,000$ samples for drawing posterior inference. Convergence and mixing are monitored through trace plots. 


\subsection{Hotspot estimation}
\label{hotspot}

{Our main goal is to exploit the observed Red Sea SST data to identify extreme hotspots, i.e., to construct a ``confidence region'' that contains joint threshold exceedances of some (very) high threshold $u$ at some future time $t_0$ with a predefined probability. Our proposed approach developed below generalizes \cite{french2013spatio} (who focus on Gaussian processes only, in the frequentist setting) and \cite{french2016credible} (extended to the Bayesian setting) to the more general and flexible case of LTP-DPMs, which better capture the joint tail behavior of complex spatiotemporal processes. While other approaches based on alternative definitions of hotspots are also possible (see, e.g., \citealp{bolin2015excursion}), the pragmatic sampling-based method of \citet{french2013spatio} that we adopt here is easy to implement, and computationally convenient as it does not rely on the solution of a complex optimization problem as in \citet{bolin2015excursion}.

Let $\mathcal{D}' = \lbrace \bm{s}_1, \ldots, \bm{s}_N \rbrace\subset \mathcal{D}$ be our discretized domain of interest. We define the exceedance region above a fixed threshold $u$ at time $t_0$ as $E_{u^+}^{0} = \lbrace \bm{s} \in \mathcal{D}' : Y_{t_0}(\bm{s}) \geq u \rbrace$. Note that because $Y_{t_0}(\cdot)$ is a random process, $E_{u^+}^{0}$ is a random set. Our goal is to find a region $\mathcal{D}^0_{u^+}$ that contains the ``true'' exceedance region $E_{u^+}^{0}$ with some predefined probability $1-\alpha$, i.e., $\textrm{Pr}(E_{u^+}^{0}\subseteq\mathcal{D}^0_{u^+}) = 1 - \alpha$. While the target region $\mathcal{D}^0_{u^+}$ is generally not unique, a valid solution to this problem may be obtained by viewing it through the lenses of multiple hypothesis testing. The proposed approach works essentially as follows. For each fixed grid cell $\bm{s}_n\in\mathcal{D}'$, we first individually test the null hypothesis $H_0: Y_{t_0}(\bm{s}_n) = u$ against the alternative $H_1: Y_{t_0}(\bm{s}_n) < u$ based on some test statistic denoted by $\tilde{Y}_{t_0}(\bm{s}_n)$. An obvious choice for $\tilde{Y}_{t_0}(\bm{s}_n)$ is to exploit (a rescaled version of) $\hat{Y}_{t_0}(\bm{s}_n)$, a predictor of $Y_{t_0}(\bm{s}_n)$. Then, to find the exceedance region, we combine these single-cell tests together by collecting all grid cells $\bm{s}_n \in \mathcal{D}'$ where we fail to reject $H_0$. However, if each test is individually performed at the confidence level $1-\alpha$, this approach accurately identifies marginal threshold exceedances, while it fails to correctly represent joint exceedances at the required \emph{overall} confidence level. Thus, we finally account for multiple testing by appropriately adjusting the critical value of the tests, in order to reach an overall family-wise error rate $\alpha$. Notice that because of spatial proximity, all the individual tests are dependent, and a sampling-based approach is here used to precisely set the critical value.




We now describe how we design the test statistic and how we set the critical value. First, consider the marginal predictive distributions based on our proposed LTP-DPM model \eqref{full_model}. We have $Y_{t_0}(\bm{s}_n) = \mu_{t_0}(\bm{s}_n) + \varepsilon_{t_0}(\bm{s}_n)$. Let $t_{01} = \lceil t_0/52 \rceil$ and $t_{02} = t_0 - 52 (t_{01} - 1)$ be the year and week, respectively, corresponding to the time point of interest, $t_0$. From \S\ref{mean_modeling} and \S\ref{overall}, the mean of  $Y_{t_0}(\bm{s}_n)$ has the form
\begin{equation} \label{prediction_mean}
    \mu_{t_0}(\bm{s}_n) = \sum_{i=1}^{2} \sum_{j=1}^{2} x^{(0;i)}_{t_{01}}  \left(\bm{x}^{(1)}_{t_{02}} \otimes \bm{x}_n^{(2;j)}\right) \bm{\beta}_{i;j},
\end{equation}
where $x_{t_{01}}^{(0;1)} = T_1^{-1/2}$, $x_{t_{01}}^{(0;2)} = (x^*_{t_{01}} - \bar{x}^*)/\{\sum_{t_1^* = 1}^{T_1} (x^*_{t_1} - \bar{x}^*)^2\}^{1/2}$ (with $\bar{x}^*$ as in \S\ref{mean_modeling}), and $\bm{x}^{(1)}_{t_{02}}$ is the $t_{02}$-th row of  $\bm{X}_1$. The density of the corresponding error term $\varepsilon_{t_0}(\bm{s}_n)$ is 
\begin{equation} \label{prediction_error}
f\left(\varepsilon\right) = \sum_{k=1}^{K} \pi_k f_{T}\left\{\varepsilon; 0, \frac{a_k - 2}{a_k} \left(\bm{h}_n \bm{\Phi}_k \bm{h}_n' + \tau_k^2 \right), a_k \right\},
\end{equation}
with $f_{T}\left(\cdot ~; 0, \tilde{\sigma}^2, \tilde{a} \right)$ the univariate Student's $t$ density function with location 0, scale $\tilde{\sigma}$, degrees of freedom $\tilde{a}$. 
Now, let $\hat{Y}_{t_0}(\bm{s}_n)$ be the posterior mean of $Y_{t_0}(\bm{s}_n)$ estimated by averaging $B$ roughly independent posterior predictive samples of $Y_{t_0}(\bm{s}_n)$ based on \eqref{prediction_mean} and \eqref{prediction_error}. Here, we have $B=10^4$, and we simulate posterior samples for the entire spatial process at once to account for spatial dependence among the tests. 
From the Bayesian central limit theorem, we then have the large-sample approximation ${\hat{Y}_{t_0}(\bm{s}_n)\mid Y_{t_0}(\bm{s}_n)} \dotsim \textrm{Normal}(Y_{t_0}(\bm{s}_n), \tilde{\sigma}_{t_0}^2(\bm{s}_n) / B)$, where $\tilde{\sigma}_{t_0}(\bm{s}_n)$ denotes the posterior standard deviation of $Y_{t_0}(\bm{s}_n)$, which we estimate from the $B$ posterior samples. 
This leads us to consider the test statistic 
\begin{equation} \label{teststat}
    \tilde{Y}_{t_0}(\bm{s}_n) = \sqrt{B}{\hat{Y}_{t_0}(\bm{s}_n) - u\over\tilde{\sigma}_{t_0}(\bm{s}_n)},
\end{equation}
and to define a rejection region for $H_0: Y_{t_0}(\bm{s}_n)=u$ (against $H_1: Y_{t_0}(\bm{s}_n) < u$) of the form $\{\tilde{Y}_{t_0}(\bm{s}_n)<C_\alpha\}$ for some critical level $C_\alpha$. Under $H_0$, $\tilde{Y}_{t_0}(\bm{s}_n)$ has an approximate standard normal distribution. We further note that this test is equivalent to testing $H_0: Y_{t_0}(\bm{s}_n)>u$ against $H_1:Y_{t_0}(\bm{s}_n)<u$.

We finally need to adjust the critical value $C_\alpha$, so that the family-wise Type I error rate is $\alpha$. This ensures a confidence level of $1 - \alpha$ for the confidence region $\mathcal{D}^0_{u^+}$. A Type I error can only occur at locations $\bm{s}_n$ within the true exceedance region, and hence to control the overall Type I error, we need to consider the test statistic values only within $E_{u^+}^0$. Therefore, the critical value $C_\alpha$ should be chosen such that 
    $\textrm{Pr} ( \min_{\bm{s_n} \in E_{u^+}^0} \{ \tilde{Y}_{t_0}(\bm{s}_n) \} < C_\alpha ) = \alpha.$ 
However, in practice, we ignore the true exceedance region $E_{u^+}^0$, and as a result the distribution of $\min_{\bm{s_n} \in E_{u^+}^0} \lbrace \tilde{Y}_{t_0}(\bm{s}_n) \rbrace$ is also unknown. To circumvent this issue, we can again exploit the $B$ posterior samples to approximate the region $E_{u^+}^0$, as well as the required probability. The critical level $C_\alpha$ is obtained as the empirical $\alpha$-quantile based on posterior samples for $\min_{\bm{s_n} \in E_{u^+}^0} \{ \tilde{Y}_{t_0}(\bm{s}_n) \}$. The steps for sampling from the posterior predictive distribution, along with estimating the critical value $C_\alpha$ empirically, and obtaining the estimated exceedance set $\mathcal{D}^0_{u^+}$, are provided in Algorithm \ref{alg1}. The proof that the family-wise error rate is indeed $\alpha$, as required for the described algorithm, is provided in the Supplementary Material.}

{ 
\begin{algorithm}
\caption{Hotspot estimation at a future time $t_0$}
\label{alg1}
{ 
\begin{small}
\begin{flushleft}
\begin{itemize}
    \item Fix an overall Type-I error $\alpha$ and a threshold level $u$.
    \linebreak
    \item Fit the model (\ref{full_model}) to the data, and obtain $B$ posterior samples from the parameters and hyperparameters.
    \linebreak
    \item For each posterior sample $b=1, \ldots, B$, 
    \begin{itemize}
        \item calculate $\mu_{t_0}(\bm{s}_n)$ according to (\ref{prediction_mean}), and let $\mu_{t_0}^{(b)}(\bm{s}_n)$ denote the the $b$-th sample.
        \item plug in $\pi_k = \pi_k^{(b)}$, and simulate $g_{t_0}^{(b)}$ from $\textrm{Pr}(g_{t_0} = k) = \pi_k^{(b)}$. Assume that $g_{t_0}^{(b)} = k_b$.
        \item plug in $a_k = a_k^{(b)}$, and simulate $\sigma_{t_0}^{2(b)}$ from $\sigma_{t_0}^2 \sim \textrm{Inverse-Gamma}\left( \frac{a_{k_b}^{(b)}}{2}, \frac{a_{k_b}^{(b)}}{2} - 1 \right)$.
        \item plug in $\bm{\Phi}_k = \bm{\Phi}_k^{(b)}$, and simulate $\bm{Z}_{t_0}^{(b)}$ from $\bm{Z}_{t_0} \sim \textrm{Normal}_L\left(\bm{0}, \bm{\Phi}_{k_b}^{(b)} \right)$.
        \item plug in $\tau_k^2 = \tau_k^{2(b)}$, and simulate $\eta_{t_0}^{(b)}(\bm{s}_n), n=1, \ldots, N$, from $\eta_{t_0}(\bm{s}_n) \overset{iid}{\sim} \textrm{Normal}\left(0, \tau_{k_b}^{2(b)} \right)$.
        \item calculate $\varepsilon_{t_0}^{(b)}(\bm{s}_n) =  \sigma_{t_0}^{(b)} \left\{\bm{h}_n' \bm{Z}_{t_0}^{(b)} + \eta_{t_0}^{(b)}(\bm{s}_n) \right\}$, the $b$-th sample from $\varepsilon_{t_0}(\bm{s}_n)$.
        \item calculate $Y_{t_0}^{(b)}(\bm{s}_n) = \mu_{t_0}^{(b)}(\bm{s}_n) + \epsilon_{t_0}^{(b)}(\bm{s}_n)$, the $b$-th posterior predictive sample from $Y_{t_0}(\bm{s}_n)$.
    \end{itemize}
    \item Based on $Y_{t_0}^{(b)}(\bm{s}_n), b=1, \ldots, B$,
    \begin{itemize}
        \item calculate $\tilde{Y}_{t_0}(\bm{s}_n)$ according to (\ref{teststat}), at each $\bm{s}_n$.
        \item identify the exceedance region $E^{b}_{u^+} = \lbrace \bm{s}_n \in \mathcal{D}' :  Y^{(b)}_{t_0}(\bm{s}_n) \geq u \rbrace$, for each $b$.
        \item calculate $\min_{\bm{s_n} \in E_{u^+}^b} \lbrace \tilde{Y}_{t_0}(\bm{s}_n) \rbrace$, for each $b$.
        \item estimate $C_\alpha$ by $\hat{C}_\alpha$, the empirical $\alpha$-quantile of $ \lbrace \min_{\bm{s_n} \in E_{u^+}^b} \lbrace \tilde{Y}_{t_0}(\bm{s}_n) \rbrace; b= 1, \ldots, B  \rbrace$.
    \end{itemize}
    \item Return $\mathcal{D}^{0}_{u^+} = \lbrace \bm{s}_n \in \mathcal{D}' :  \tilde{Y}_{t_0}(\bm{s}_n) \geq \hat{C}_\alpha \rbrace $.
\end{itemize}
\end{flushleft}
\end{small}
}  
\end{algorithm}
} 

\section{{ Data application}}
\label{application}




\subsection{Model fitting and cross-validation study}
\label{crossvalidation}


{Here we present the results based on RCP 8.5, considering it as a relatively pessimistic greenhouse gas emission scenario. The results based on RCP 4.5 (moderate mitigation pathway) are provided in the Supplementary Material.} To fit our model, we first need to fix the number of mixture components $K$, and the number of spatial basis functions (EOFs), $L$. Both $K$ and $L$ affect the bias-variance trade-off and the computational burden. We specify $K=5,10$, and $L=\arg\max_{l\in\{1,\ldots,N\}}\{\lambda_l \geq q \lambda_1\}$ with $q=0.005,0.01$, where $\lambda_1 > \lambda_2 > \ldots, \lambda_N$ are the ordered eigenvalues of $\hat{\bm{\Sigma}}$ (recall \S\ref{spatdep}). For $q=0.01$ and $q=0.005$, we denote the values of $L$ by $L_1$ and $L_2$, respectively. For the Red Sea SST data, we obtain $L_1 = 15$ and $L_2 = 24$ for all the seven weekly sub-datasets. We then compare the different choices of $K$ and $L$ by cross-validation. Moreover, we also compare our proposed LTP-DPM model to some simpler parametric and semiparametric sub-models, namely LGP ($K=1$, $a_k = \infty$), LTP ($K=1$) and LGP-DPM ($a_k = \infty$).


For model comparison, we divide each weekly Red Sea SST sub-dataset into two parts, using the years 1985--2010 (1352 weeks) for training and keeping the years 2011--2015 (260 weeks) for testing. For each spatiotemporal observation in the test set, we estimate the posterior predictive distribution based on the posterior samples and confront the estimated distribution with the test observation. Because our primary goal is to predict high threshold exceedances, we use the Brier score (BS) and the tail-weighted continuous rank probability score (TWCRPS), proposed by \cite{gneiting2007strictly} and \cite{gneiting2011comparing}, respectively. For a single test sample $y$, the BS at a given level $u$ is defined as ${\rm BS}_u(y,F)=\{\mathbb{I}_{\lbrace y > u \rbrace} - \bar{F}(u)\}^2$, where $\bar{F}(\cdot)=1-F(\cdot)$ is the survival function corresponding to the posterior predictive distribution $F$. For a single test sample $y$, the TWCRPS is defined as ${\rm TWCRPS}_w(y,F)=\int_{-\infty}^{\infty} w(x) \left\{F(x) - \mathbb{I}_{\lbrace y \leq x \rbrace} \right\}^2 {\rm d}x$, where $w(\cdot)$ is a non-negative weight function. To focus on the upper tail, we use $w(x) = \mathbb{I}_{\lbrace x > u \rbrace}$ so that ${\rm TWCRPS}_w(y,F)=\int_{u}^{\infty}{\rm BS}_x(y,F){\rm d}x$. Then, we define the Brier skill score (BSS) and tail-weighted continuous rank probability skill score (TWCRPSS) for a model $M$ as 
\begin{equation}
   \textrm{BSS}_{M}  = {\textrm{BS}_{\textrm{LGP}} - \textrm{BS}_{M}\over\textrm{BS}_{\textrm{LGP}}} \times 100 \%,  \textrm{TWCRPSS}_{M}  = {\textrm{TWCRPS}_{\textrm{LGP}} - \textrm{TWCRPS}_{M}\over\textrm{TWCRPS}_{\textrm{LGP}}} \times 100 \%,
\end{equation}
{using LGP with $L=L_1$ as the benchmark, where $\textrm{BS}_{M}$ and $\textrm{TWCRPS}_{M}$ are the short-hand notation for the BS and TWCRPS for a model $M$, respectively. Higher values of BSS or TWCRPSS indicate better prediction performance. We report the results by averaging values over the test set. We chose $L_1$ basis functions for LGP benchmark because it corresponds to the simplest (i.e., most parsimonious) model fitted here. Figure~\ref{crps_comparison} reports the skill scores plotted as a function of the threshold $u$, which ranges between the $95\%$ and the $99.9\%$-quantiles of the SST data. 
\begin{figure}[t!]
\begin{center}
\adjincludegraphics[width = \linewidth, trim = {{.12\width} {.50\width} {.12\width} {.46\width}}, clip]{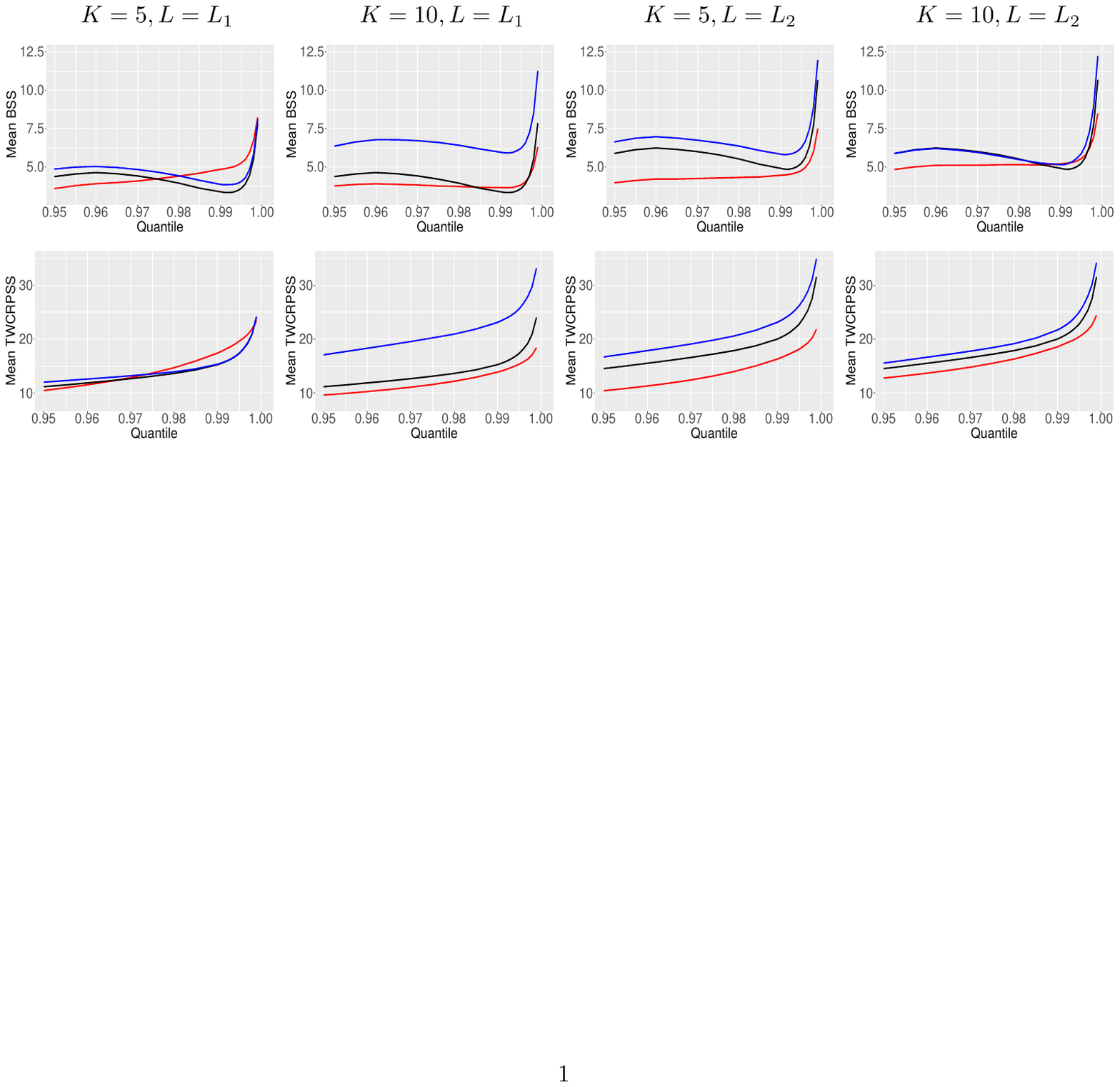}
\end{center}
\caption{Averaged BSS (top) and TWCRPSS (bottom) for the models LTP (black), LGP-DPM (red) and LTP-DPM (blue) considering the LGP as the reference model, plotted as a function of the threshold $u$ ranging from the $95\%$-quantile to the $99.9\%$-quantile of the observed data. Higher values of BSS and TWCRPSS indicate better prediction performance.}
	\label{crps_comparison}
\end{figure}
As all scores are positive, the benchmark LGP model is consistently the worst, while our proposed LTP-DPM model (blue curves) performs better than the others in most of the cases. Comparing across the choices of $K$ and $L$, the predictive performances of the LTP-DPM model for the cases $K=10, L = L_1$ and $K=5, L = L_2$ are comparable, while the skill scores are noticeably lower for $K=5, L = L_1$ (most parsimonious LTP-DPM) and slightly lower for $K=10,L=L_2$ (most complex LTP-DPM). Among the two intermediate cases with best performances ($K=10, L = L_1$ and $K=5, L = L_2$), the model with $K=10, L = L_1$ is the most parsimonious. As there is no visible difference in their performances, we therefore proceed with $K=10, L = L_1$ to draw inference.}



{Detailed goodness-of-fit diagnostics for the LTP-DPM model are provided in the Supplementary Material. To summarize, the mean and standard deviation profiles are very well estimated overall, as demonstrated by very small root mean squared differences between their corresponding empirical and fitted model-based estimates. Moreover, while the difference between empirical and fitted pairwise correlations varies between -0.1 and 0.1, the empirical and the fitted $\chi$-coefficients appear to be more variable. There is indeed a higher degree of uncertainty involved in the empirical $\chi$-coefficients due to the sparsity of extreme events, which results in larger differences with their model-based counterparts. This, however, does not necessarily indicate a lack of fit, as further explained in the Supplementary Material.}







\begin{figure}[t!]
\centering
	\adjincludegraphics[height = 0.19\linewidth, width = 0.19\linewidth, trim = {{.0\width} {.0\width} {.0\width} {.05\width}}, clip]{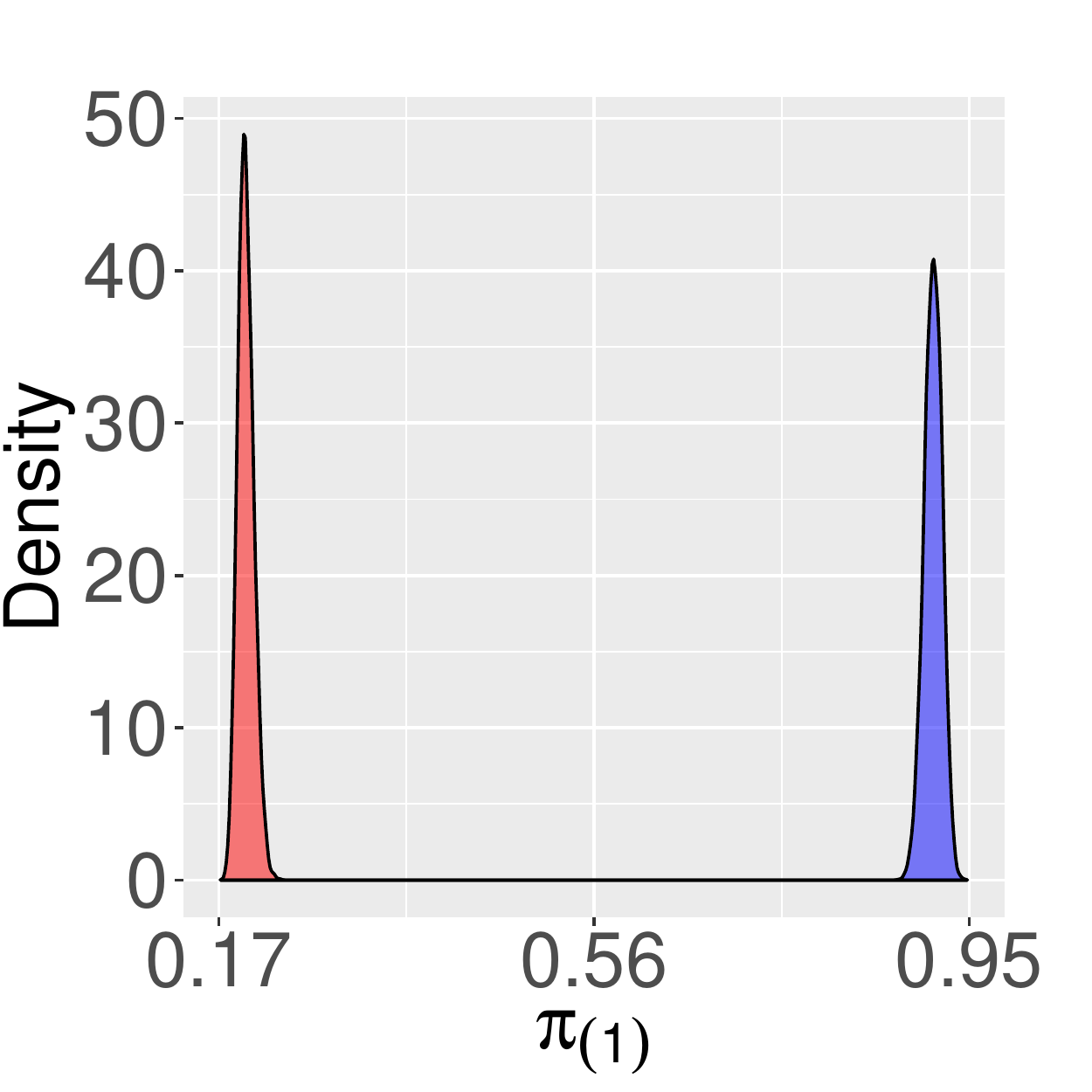}
	\adjincludegraphics[height = 0.19\linewidth, width = 0.19\linewidth, trim = {{.0\width} {.0\width} {.0\width} {.05\width}}, clip]{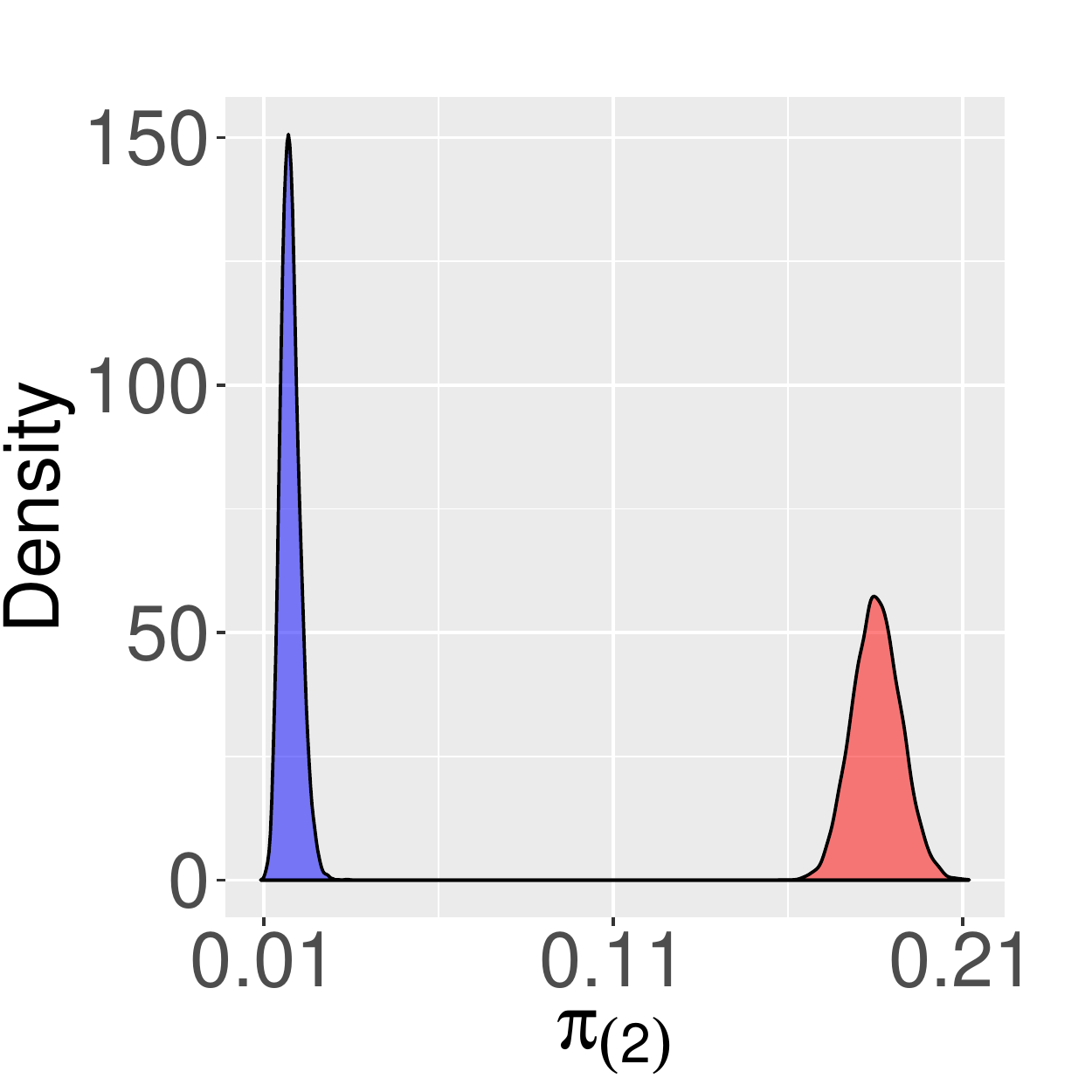}
	\adjincludegraphics[height = 0.19\linewidth, width = 0.19\linewidth, trim = {{.0\width} {.0\width} {.0\width} {.05\width}}, clip]{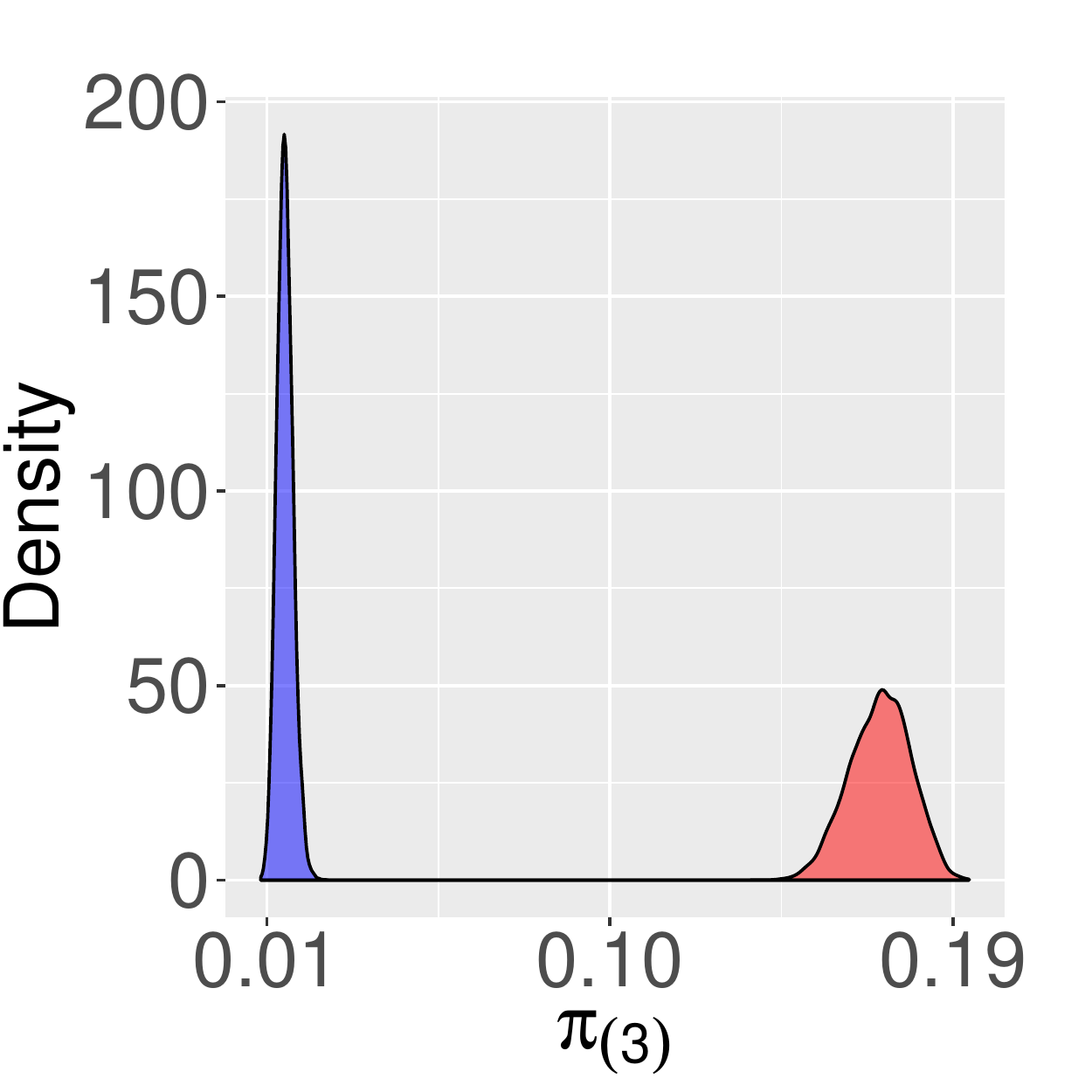}
	\adjincludegraphics[height = 0.19\linewidth, width = 0.19\linewidth, trim = {{.0\width} {.0\width} {.0\width} {.05\width}}, clip]{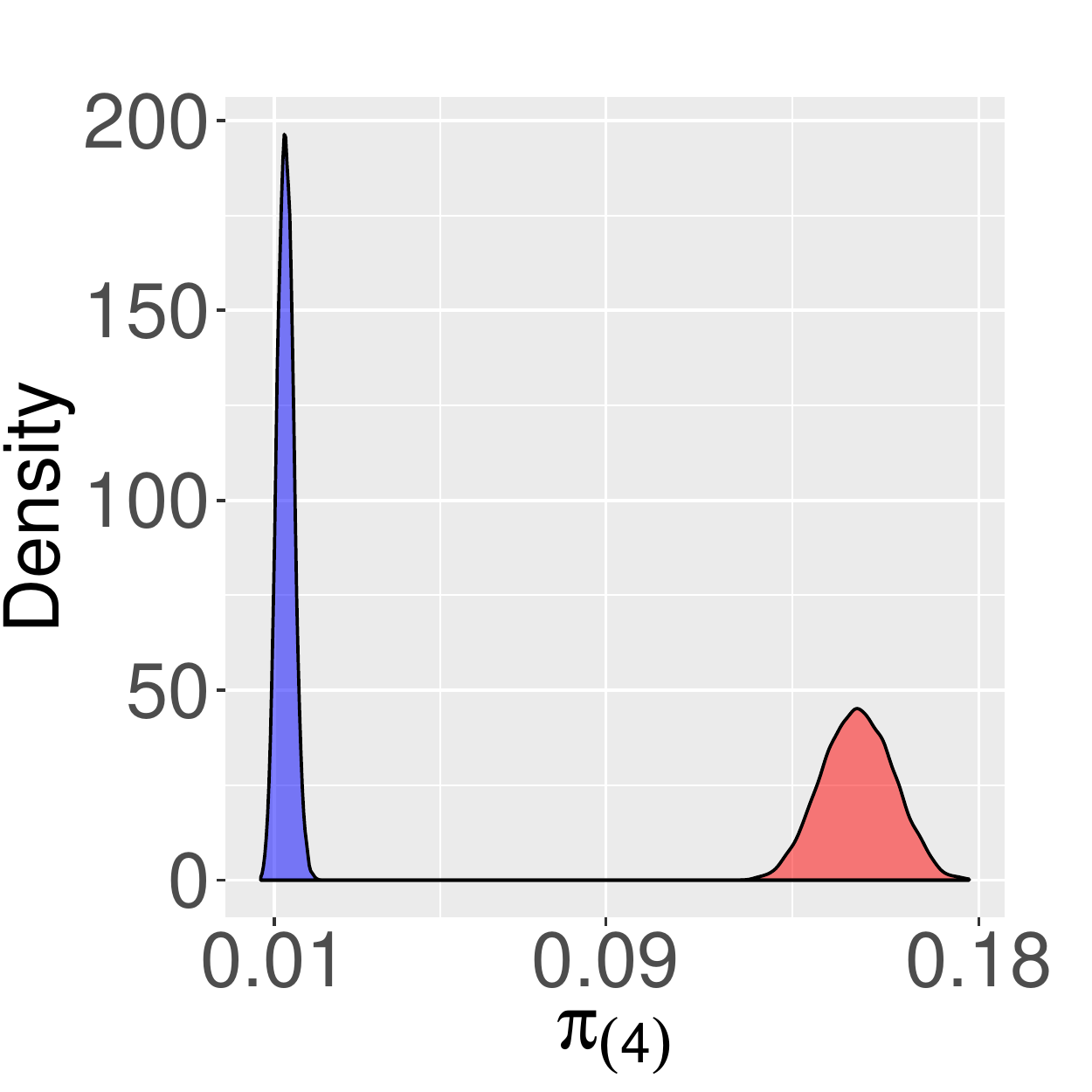}
	\adjincludegraphics[height = 0.19\linewidth, width = 0.19\linewidth, trim = {{.0\width} {.0\width} {.0\width} {.05\width}}, clip]{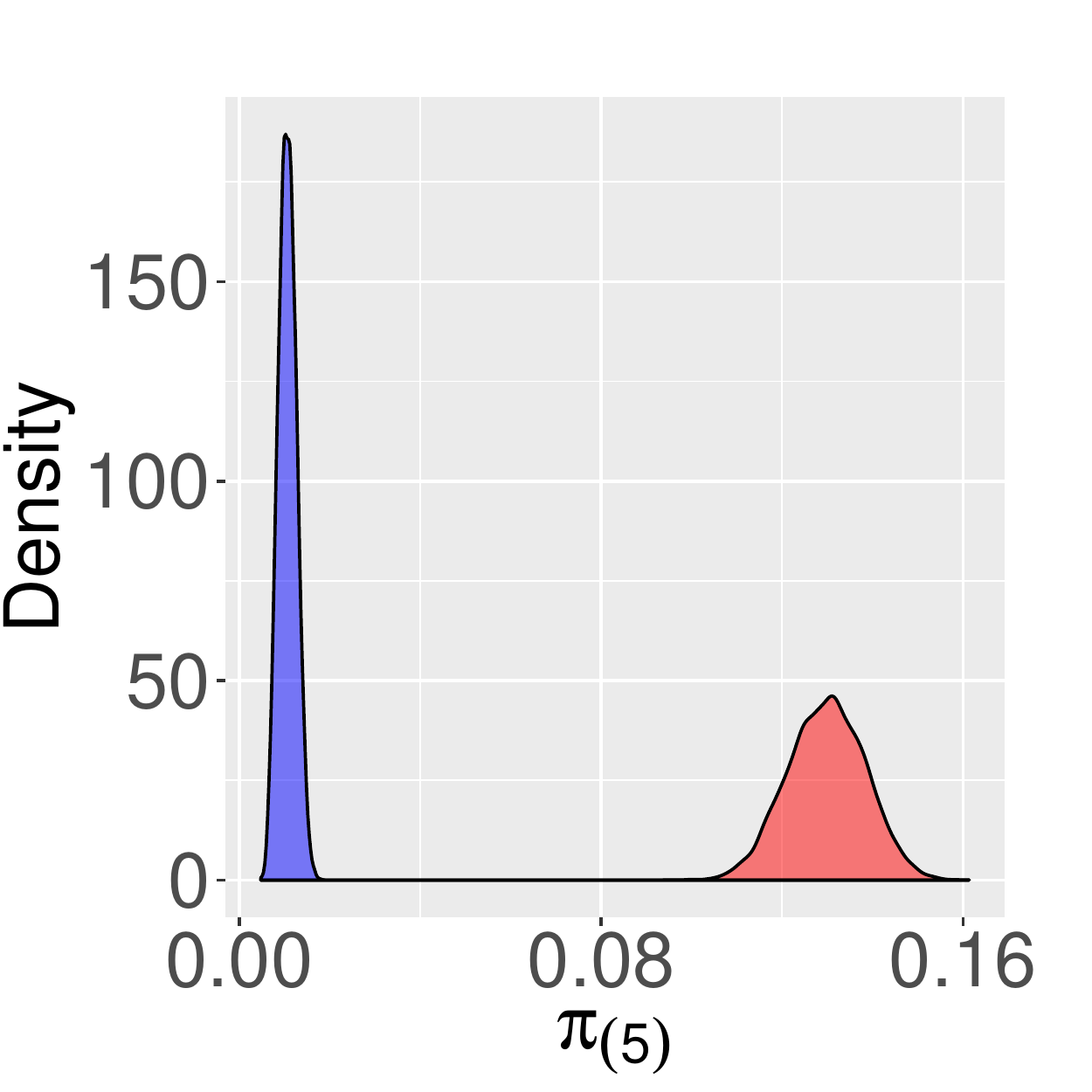}
	\adjincludegraphics[height = 0.19\linewidth, width = 0.19\linewidth, trim = {{.0\width} {.0\width} {.0\width} {.05\width}}, clip]{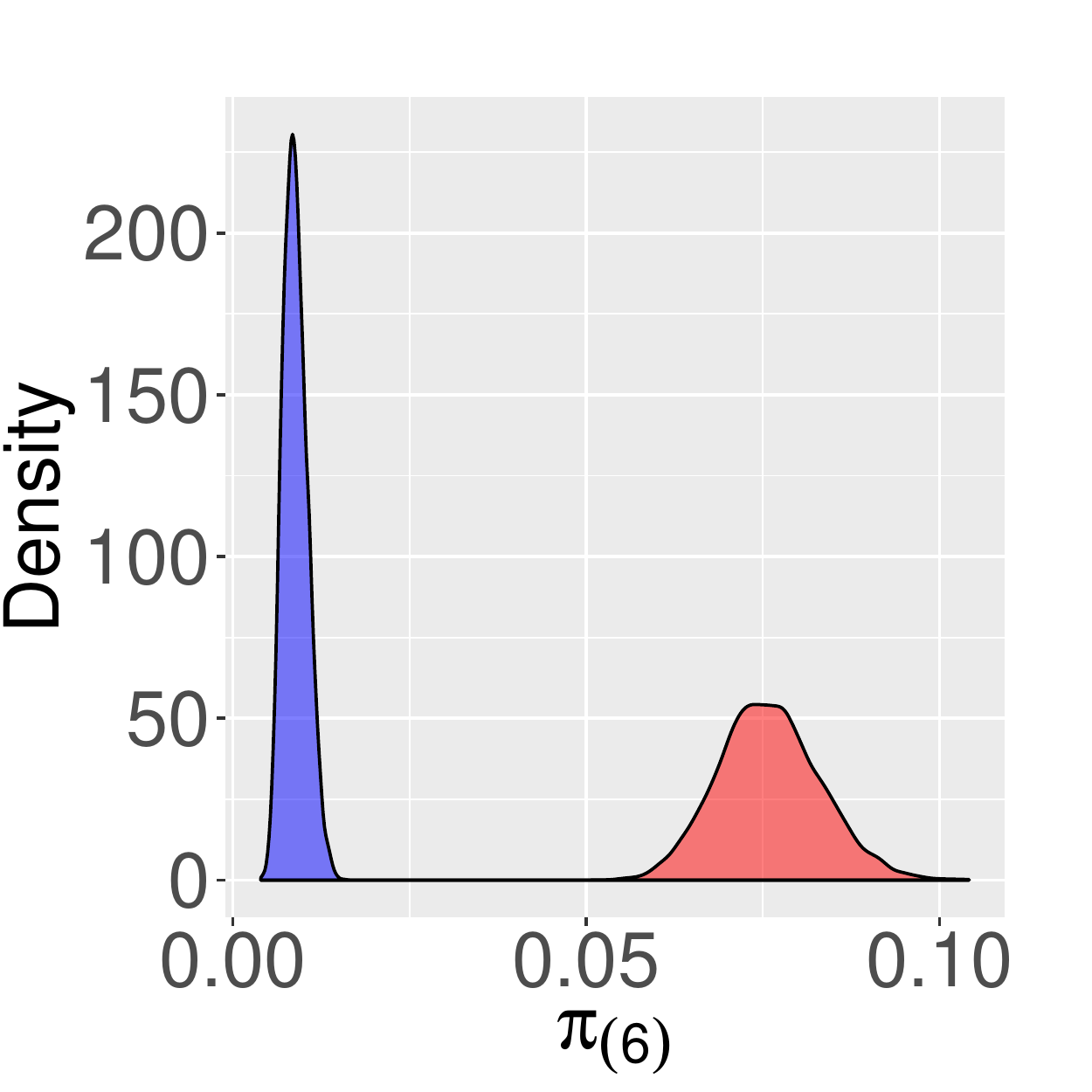}
	\adjincludegraphics[height = 0.19\linewidth, width = 0.19\linewidth, trim = {{.0\width} {.0\width} {.0\width} {.05\width}}, clip]{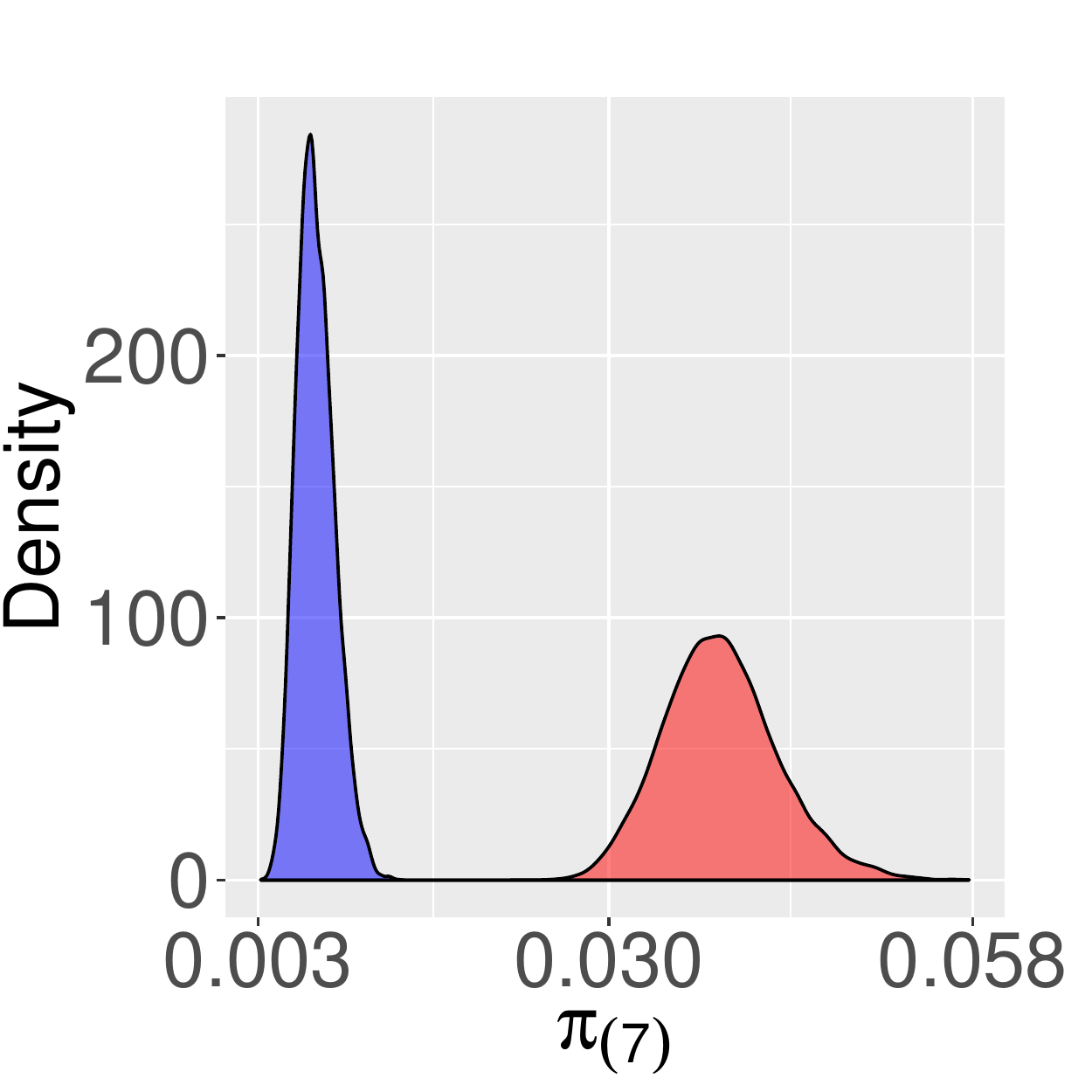}
	\adjincludegraphics[height = 0.19\linewidth, width = 0.19\linewidth, trim = {{.0\width} {.0\width} {.0\width} {.05\width}}, clip]{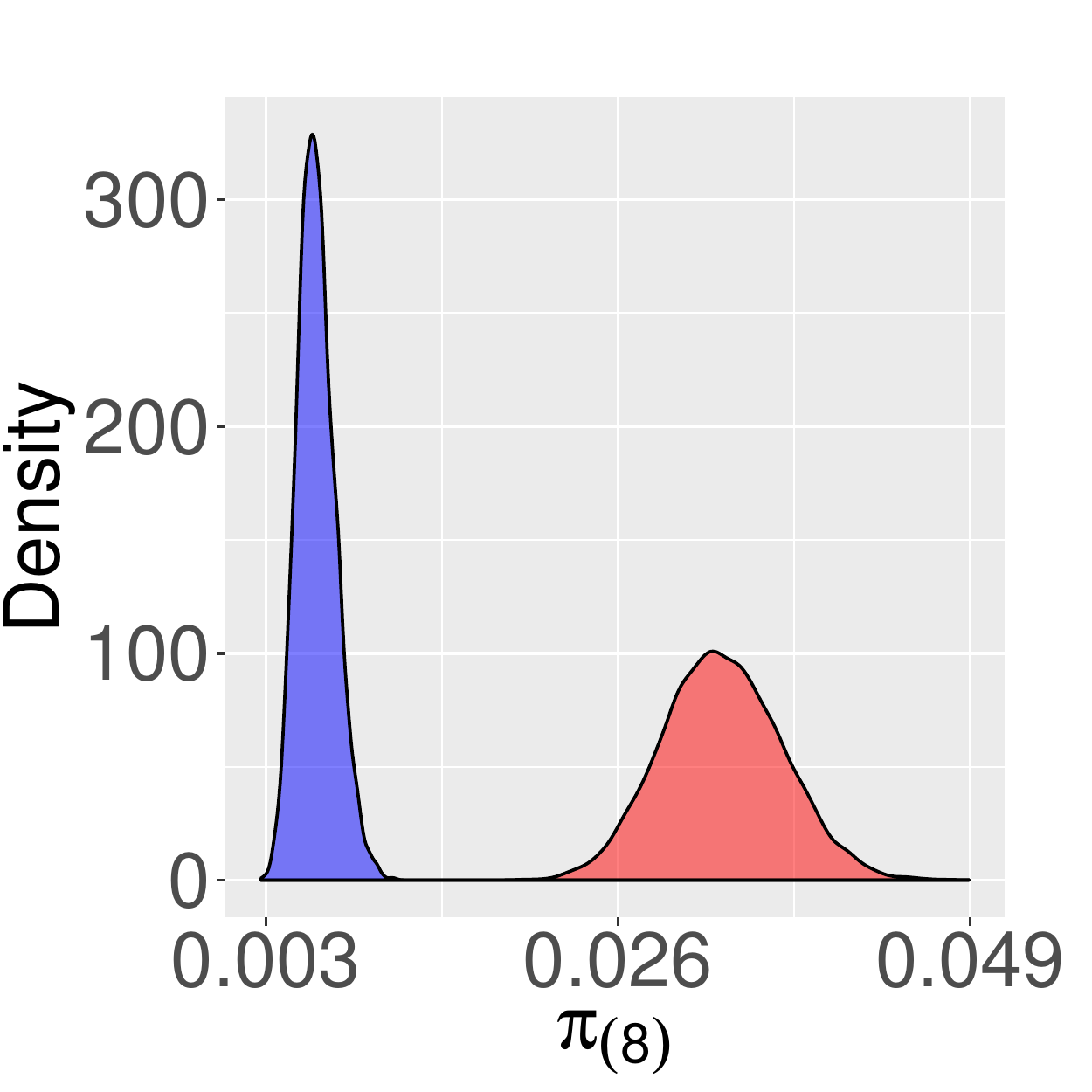}
	\adjincludegraphics[height = 0.19\linewidth, width = 0.19\linewidth, trim = {{.0\width} {.0\width} {.0\width} {.05\width}}, clip]{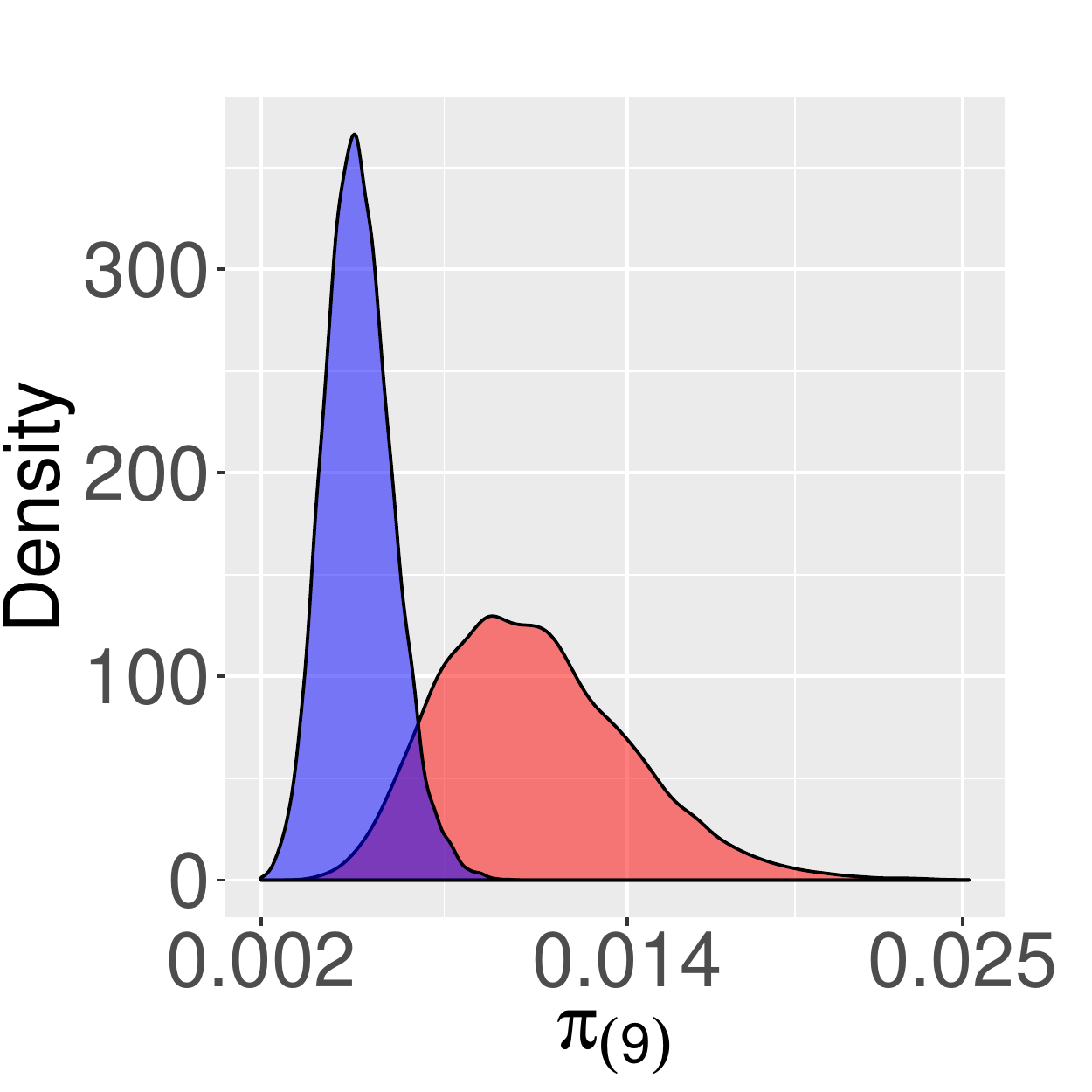}
	\adjincludegraphics[height = 0.19\linewidth, width = 0.19\linewidth, trim = {{.0\width} {.0\width} {.0\width} {.05\width}}, clip]{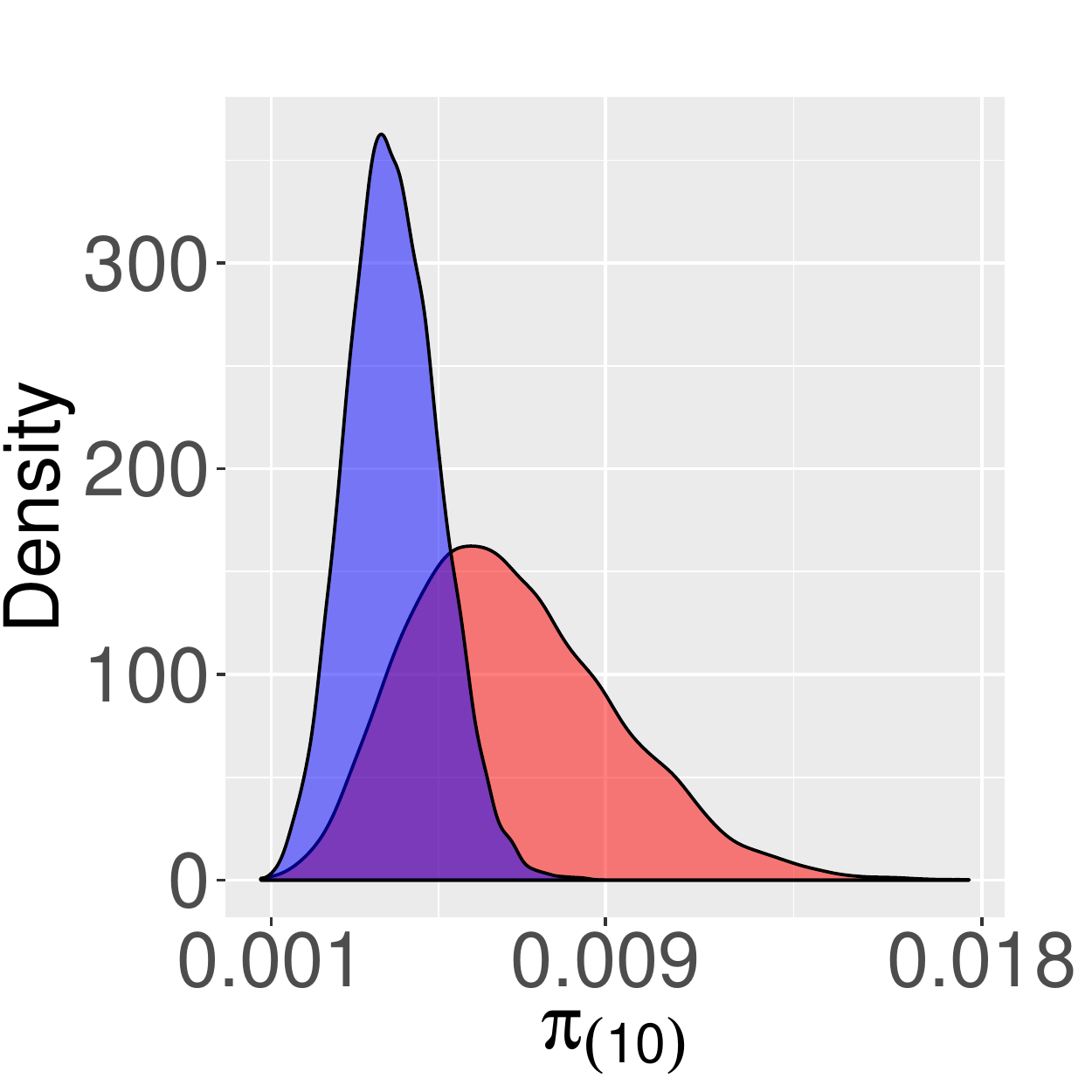}
	\caption{Posterior densities of the ten ordered stick-breaking probabilities $\pi_{(1)}>\cdots>\pi_{(K)}$, based on fitting the LGP-DPM (red) and LTP-DPM (blue) models with $K=10$ mixture components and $L=L_1$ spatial basis functions.}
	\label{posterior_pi}
\end{figure}

To explore the DPM models further, we investigate the posterior distribution of stick-breaking probabilities $\pi_k$. Because the $\pi_k$'s are not identifiable due to label-switching within the MCMC algorithm, Figure~\ref{posterior_pi} displays the estimated densities of the ordered probabilities, $\pi_{(k)}$, $k=1, \ldots, K$, for $K=10$ and $L=L_1$. For the LTP-DPM model, one single cluster fits a large proportion of the temporal replicates, while the remaining nine clusters capture the behavior of ``abnormal weeks'', i.e., extreme events. This property is desirable as it allows us to have a separate control over extreme events, being described by different mixture-specific parameters. On the other hand, the LGP-DPM model allocates large probabilities to many clusters. The thin-tailed LGP cluster components fail to allocate the bulk of the distribution into a single cluster for our heavy-tailed SST data. This makes the identification of the bulk and the tail more challenging and requires significantly more clusters for model fitting.



\subsection{Estimated time trend and return levels}



We now discuss the estimated spatial maps of the decadal rate of change (DRC) in mean SST and the return levels (adjusted for non-stationarity) based on our best LTP-DPM model.

\begin{figure}[t!]
\begin{center}
    	\adjincludegraphics[height = 0.4\linewidth, width = 0.26\linewidth, trim = {{.05\width} {.0\width} {.32\width} {.0\width}}, clip]{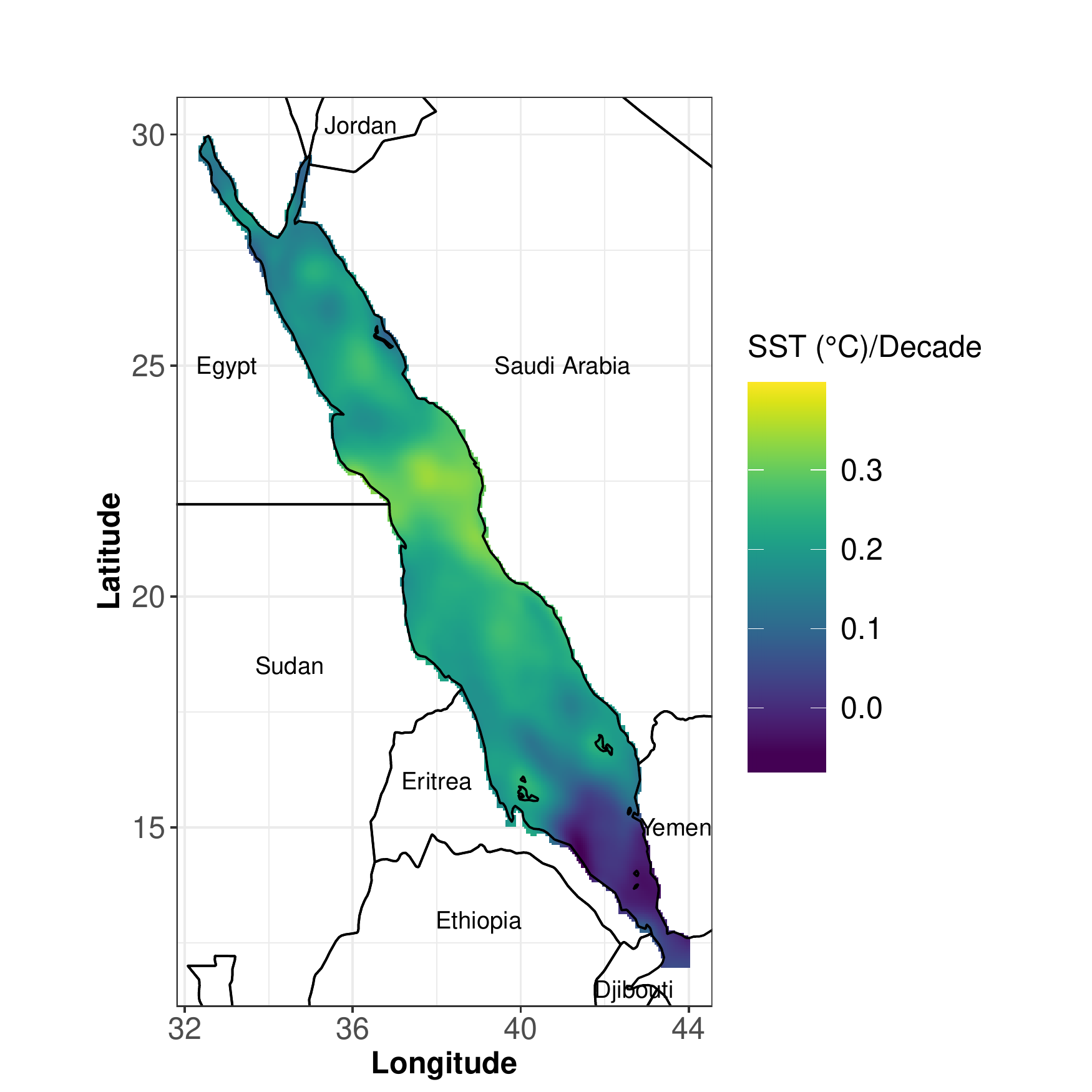}
	\adjincludegraphics[height = 0.4\linewidth, width = 0.26\linewidth, trim = {{.05\width} {.0\width} {.32\width} {.0\width}}, clip]{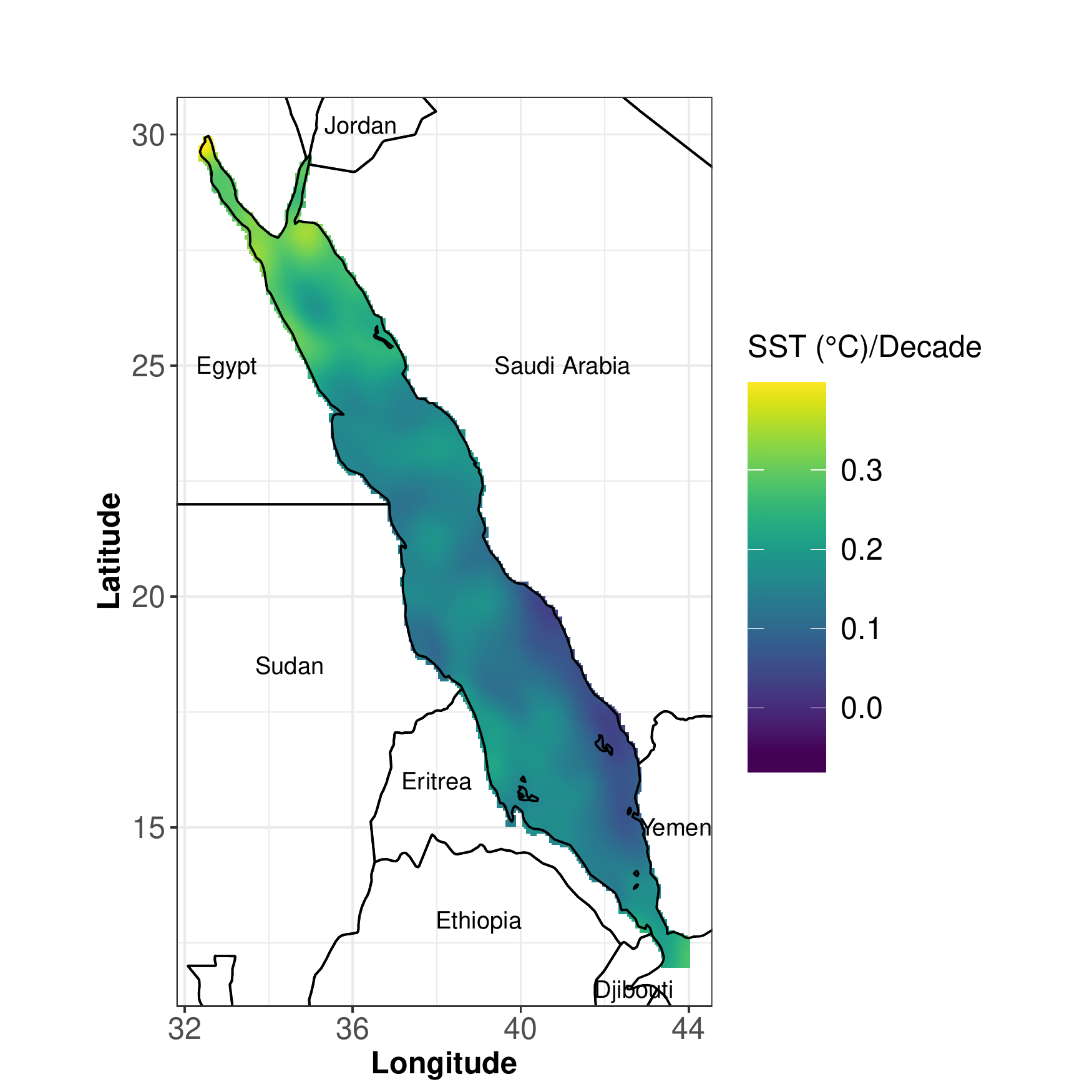}
	\adjincludegraphics[height = 0.4\linewidth, width = 0.38\linewidth, trim = {{.05\width} {.0\width} {.0\width} {.0\width}}, clip]{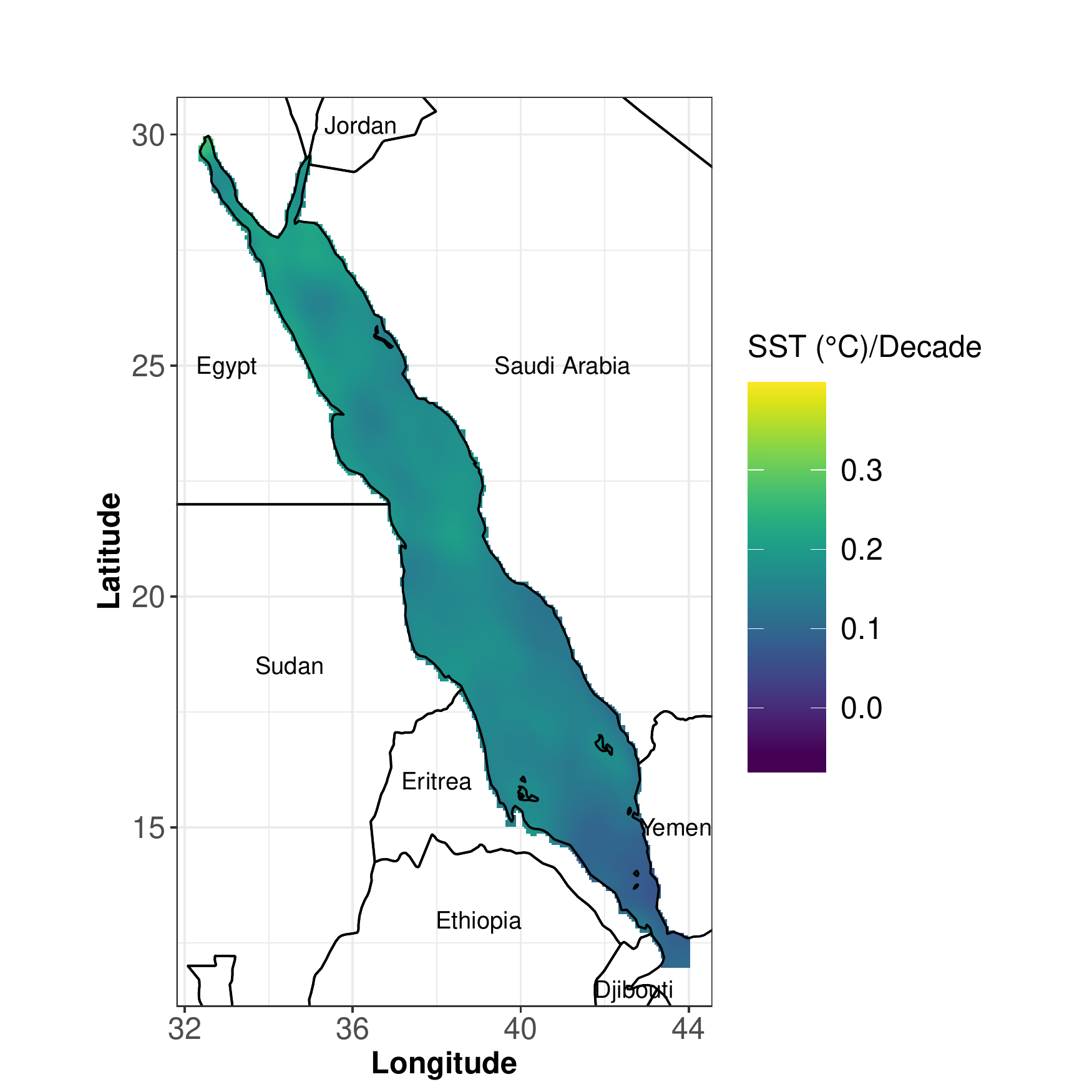}
\end{center}
\caption{Estimated decadal rate of change in mean SST profile for the coolest (7-th, left) and the warmest (34-th, middle) weeks along with the estimated overall rate obtained by averaging across the 52 weeks (right). The sub-figures are on the same scale.}
\label{three_roc}
\end{figure}

{As the mean SST incorporates the simulated RCP-based SST projections as covariate, we define the DRC for a specific week $t_2$ at a spatial location $\bm{s}$ as $\textrm{DRC}_{t_2}(\bm{s}) = 10\{\mu(T_1, t_2, \bm{s}) - \mu(1, t_2, \bm{s})\} / (T_1 - 1)$, where the notation is as described in \S \ref{mean_modeling}.} We present in Figure \ref{three_roc} the spatial maps of the estimated weekly-varying DRC in mean SST across the Red Sea for Weeks 7 and 34 (the coolest and warmest weeks based on the averaged observed SST, respectively) along with the overall DRC obtained by averaging across the 52 weeks. The spatial patterns of DRC for the two weeks are significantly different and broadly consistent with the exploratory analysis (Figure \ref{slopes}). For Week 7 (winter), the highest DRC values are near the latitude 22$^\circ$N while the lowest values are observed near the southern end of the Red Sea. For Week 34 (summer), the highest DRC values are near the northern tip of the Gulf of Suez and over a large coastal region of Egypt as well as the coastal region of northwest Saudi Arabia while the lowest values are observed near the coastal region of the southwest Saudi Arabia. The spatial map of the overall DRC is smoother than the weekly profiles, with higher values being observed near the coastal region of Egypt. In addition, we also calculate the corresponding $t$-statistic for each spatial location; see the Supplementary Material. A value of $|t|> 2$ indicates a (site-wise) significant change at the 95\% confidence level in mean SST over the years 1985--2015. While negative estimates of DRC are obtained near the coast of Eritrea for Week 7 (for a total 896 grid cells), all such DRC values are not significant except for 35 grid cells. Considering the overall DRC, the $t$-statistics vary between 5.92 and 39.18 indicating that the overall DRC is positive and highly significant at all the grid cells. 


A $T_0$-year return level of a stationary weekly time series is (approximately) the $(1-1/[52T_0])$-th quantile of its marginal distribution, given that there are about 52 weeks per year. The LTP-DPM model, however, has a nonstationary mean and involves both trend and seasonality. Thus, quantiles---hence return levels---change over time. As the model involves a trend that is linearly related to the RCP 8.5-based SST estimates for each week, we follow \cite{cheng2014non} and consider the average RCP 8.5-based SST estimate within the return period, and calculate the mean SST and return level accordingly. Here, we concentrate specifically on some of the warmest weeks. While the SST spatial average is maximum for Week 34, Week 40 is the hottest week for 2860 grid cells spread between Eritrea and southwest Saudi Arabia, which covers some major coral reefs. The results for Weeks 33 and 34 are discussed in the Supplementary Material. For the Week 40, Figure \ref{three_return_levels} provides the spatial maps of 10-year, 20-year and 50-year estimated return levels considering 2020 as the reference year. For all three return periods, estimates are lower near the Gulf of Suez and the Gulf of Aqaba, and higher across a large region between the coast of Eritrea and the southwest of Saudi Arabia. 
This suggests that very extreme sea temperatures might occur over the next century, potentially causing severe ecological damages to endemic species (e.g., certain types of corals), and may have also economic consequences for neighboring countries.

\begin{figure}[t!]
\begin{center}
    	\adjincludegraphics[height = 0.4\linewidth, width = 0.25\linewidth, trim = {{.13\width} {.0\width} {.29\width} {.0\width}}, clip]{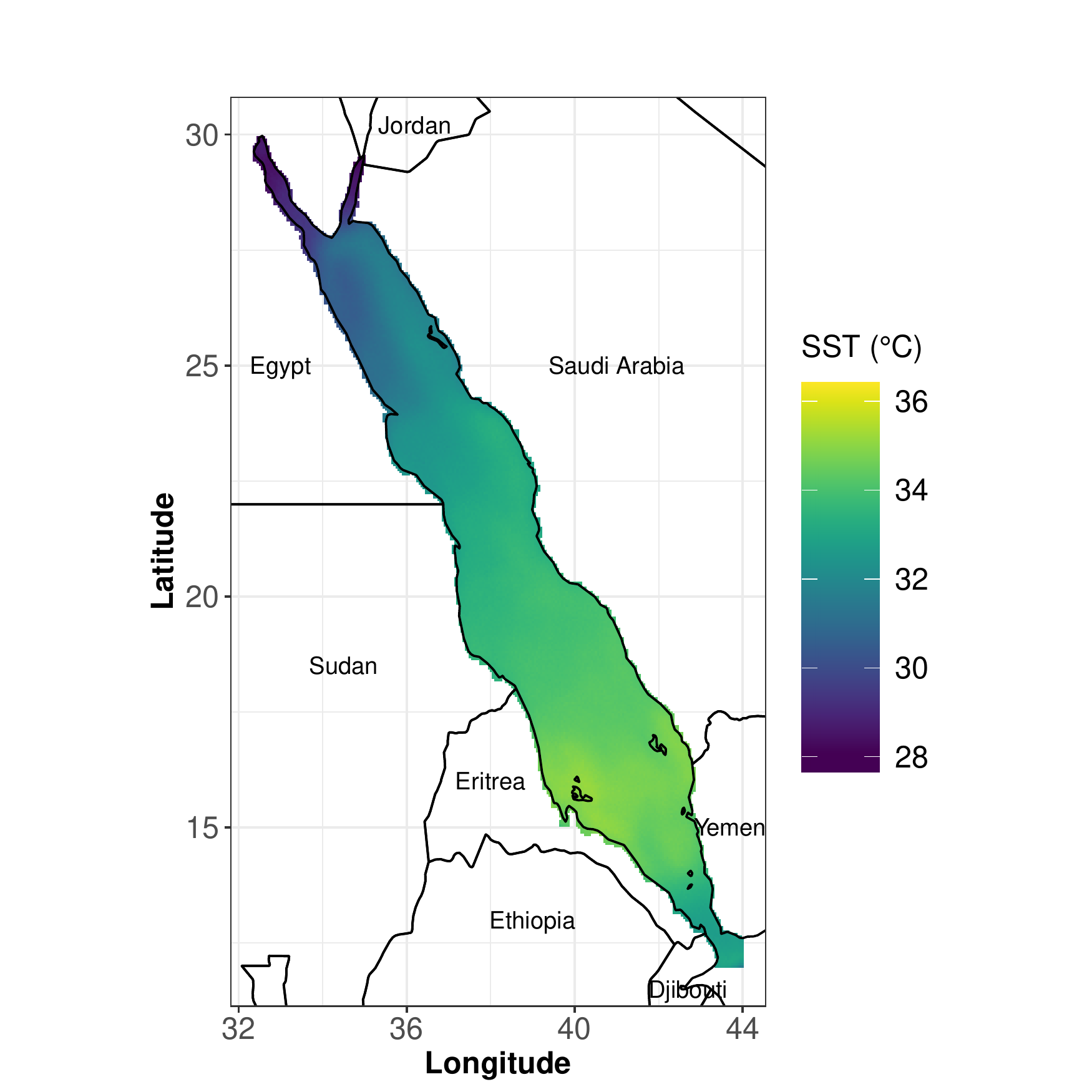}
	\adjincludegraphics[height = 0.4\linewidth, width = 0.25\linewidth, trim = {{.13\width} {.0\width} {.29\width} {.0\width}}, clip]{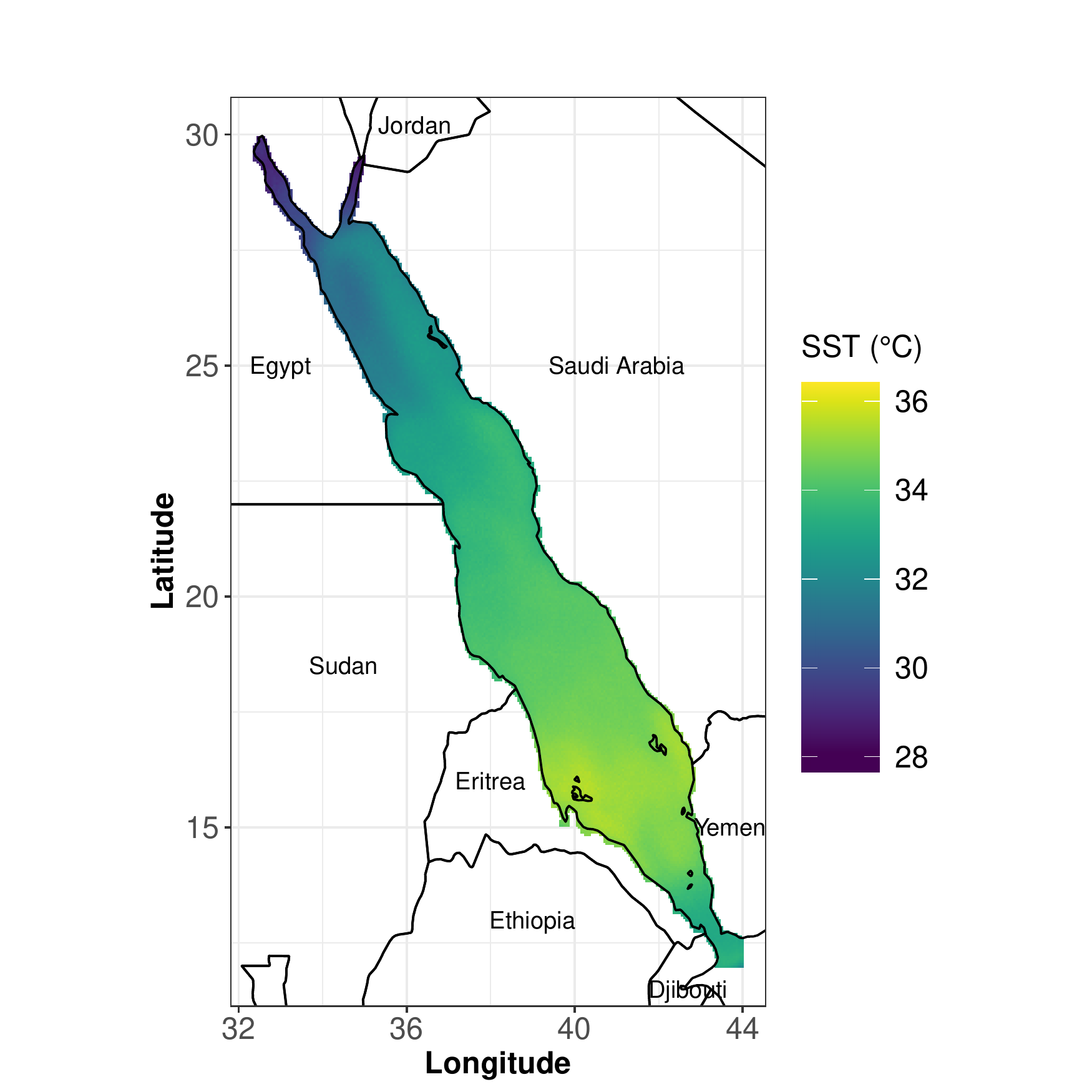}
	\adjincludegraphics[height = 0.4\linewidth, width = 0.32\linewidth, trim = {{.13\width} {.0\width} {.10\width} {.0\width}}, clip]{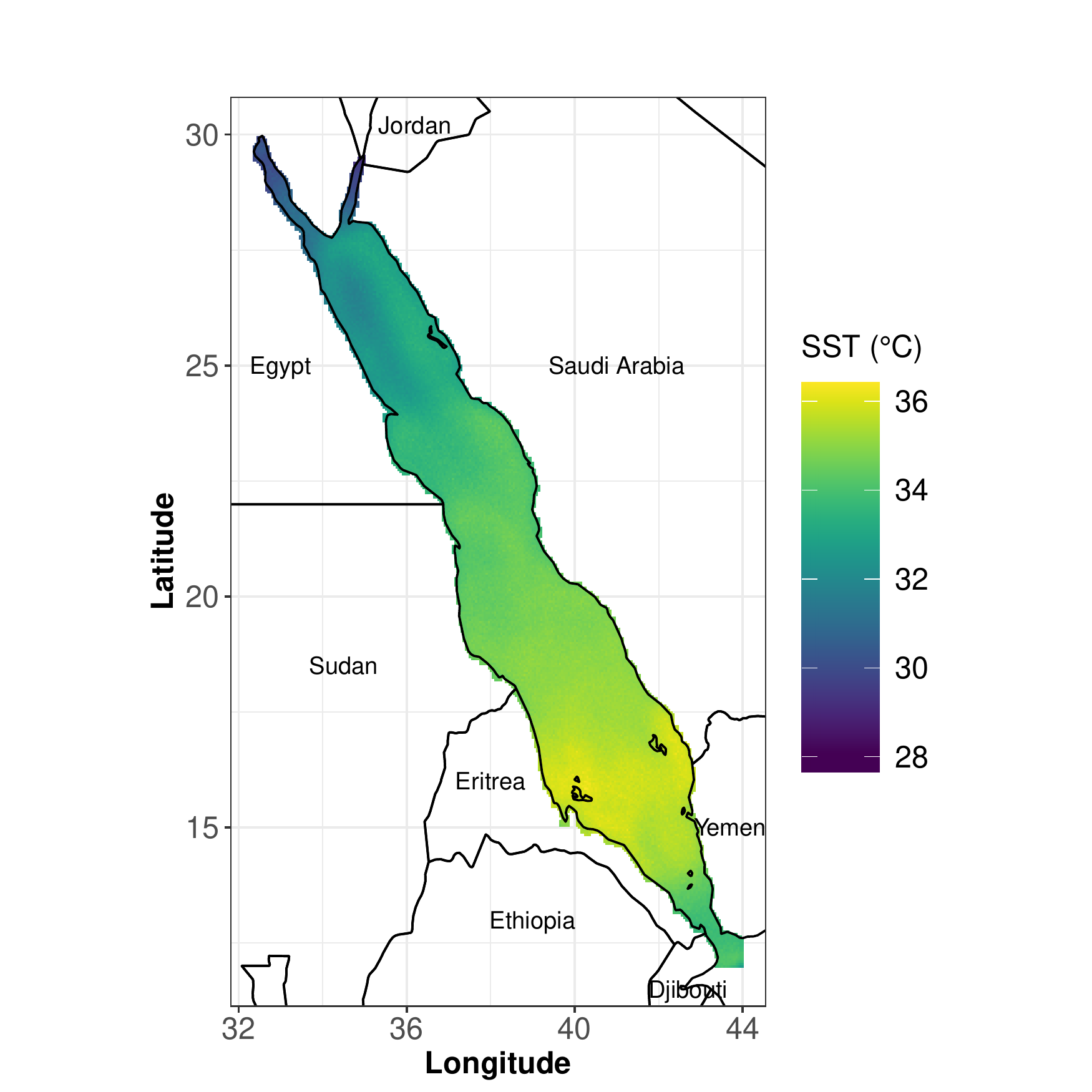}
\end{center}
\caption{Estimated 10 (left), 20 (middle) and 50 (right)-year return levels for Week 40, taking 2020 as the reference year, and following \cite{cheng2014non} to account for non-stationarity. All sub-figures share the same scale.}
\label{three_return_levels}
\end{figure}


\subsection{Estimated joint exceedance probabilities}
\label{estimated_exceedprobs}

\begin{figure}[t!]
\centering
\adjincludegraphics[height = 0.6\linewidth, width = 0.95\linewidth, trim = {{.08\width} {.40\width} {.12\width} {.39\width}}, clip]{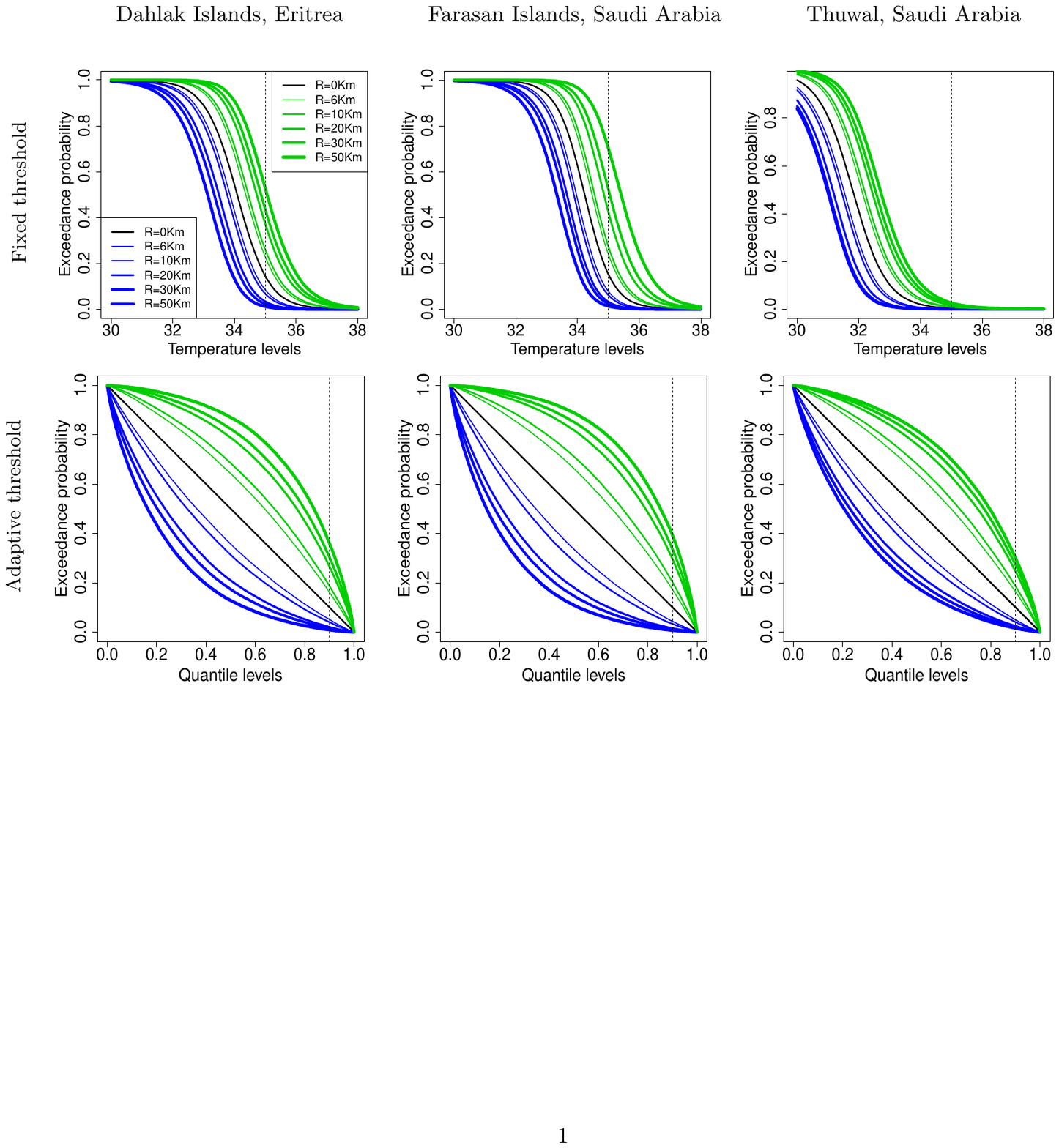}
\caption{Exceedance probabilities of SST across all the grid cells (blue) or at least one grid cell (green) within 0 km (the marginal case), 6 km (including the first order neighbors), 10 km, 20 km, 30 km and 50 km distances from (40.13$^\circ$E, 15.77$^\circ$N) near Dahlak Islands of Eritrea, (41.88$^\circ$E, 16.82$^\circ$N) near Farasan Islands of Saudi Arabia and (38.88$^\circ$E, 22.37$^\circ$N) near Thuwal of Saudi Arabia at different temperature levels for Week 40 of the reference year 2099 (corresponds to the highest RCP 8.5-based SST projection until 2100), 
considering fixed (top) or spatially-varying (bottom) temperature thresholds. Vertical dashed lines show reference thresholds corresponding to $u=35^\circ$C (top) and the $p=0.9$-quantile (bottom).}
\label{temperature_exceedance}
\end{figure}

We now investigate model-based estimates of joint exceedance probabilities in more detail, and consider three specific regions: the Dahlak Islands of Eritrea (40.13$^\circ$E, 15.77$^\circ$N), the Farasan Islands of Saudi Arabia (41.88$^\circ$E, 16.82$^\circ$N), and the region of Thuwal in Saudi Arabia (38.88$^\circ$E, 22.37$^\circ$N), where large coral reefs are present. Considering the RCP 8.5-based Red Sea SST projections until 2100, the highest projection (33.85$^\circ$C) corresponds to the year 2099; see Figure~\ref{fig:RCP}. For this particular year, we estimate two types of exceedance probabilities, namely $\textrm{Pr}\left( \cup_{\bm{s}_n \in \mathcal{D}_0} \lbrace Y_{t_0}(\bm{s}_n) > u  \rbrace \right)$ and $\textrm{Pr}\left( \cap_{\bm{s}_n \in \mathcal{D}_0} \lbrace Y_{t_0}(\bm{s}_n) > u  \rbrace \right)$, for a range of high temperature values $u$ and different neighborhoods $\mathcal{D}_0$, i.e., sets of grid cells within a certain distance from the three specific locations. For $\mathcal{D}_0$, we consider distances 0 km (for which both types of probabilities coincide with the marginal exceedance probability), 6 km (including the first order neighbors), 10 km, 20 km, 30 km and 50 km. A large value of $\textrm{Pr}\left( \cup_{\bm{s}_n \in \mathcal{D}_0} \lbrace Y_{t_0}(\bm{s}_n) > u  \rbrace \right)$ indicates a high probability that \emph{at least one} grid cell within $\mathcal{D}_0$ exceeds the threshold $u$. 
A large value of $\textrm{Pr}\left( \cap_{\bm{s}_n \in \mathcal{D}_0} \lbrace Y_{t_0}(\bm{s}_n) > u  \rbrace \right)$ indicates a high probability that \emph{all} grid cells within $\mathcal{D}_0$ exceed the threshold $u$. 
These probabilities are different across weeks. The results for Weeks 33 and 34 are provided in the Supplementary Material. For Week 40, the values are presented in the top panel of Figure \ref{temperature_exceedance}. For Dahlak Islands of Eritrea and Farasan Islands of Saudi Arabia, the exceedance probabilities are high even for high thresholds. As Week 40 is a post-summer week in the northern Red Sea, the exceedance probabilities are comparatively lower for Thuwal, Saudi Arabia.
At a threshold of 35$^\circ$C, the estimated probability values (combining both types of exceedance probabilities) range within $(0.0122, 0.5277)$, $(0.0137, 0.7031)$ and $(0.0004, 0.0264)$ for the three regions, respectively.

We also estimate the same exceedance probabilities, but instead of calculating them over a range of spatially-constant temperature levels, we use an adaptive site-specific threshold. Fixed or adaptive thresholds might be useful in different contexts: while a fixed threshold is directly interpretable on the temperature scale, a spatially-varying threshold is more in line with coral bleaching theory \citep[see, e.g.,][]{genevier2019marine}. 
For any $p \in (0, 1)$, let $Q^{(n)}_{t_0}(p)$ be the $p$-th quantile function of the distribution of $Y_{t_0}(\bm{s}_n)$ at the $n$-th grid cell, $\bm{s}_n$, for the reference time $t_0$. We estimate the probabilities $\textrm{Pr}( \cup_{\bm{s}_n \in \mathcal{D}_0} \lbrace Y_{t_0}(\bm{s}_n) > Q^{(n)}_{t_0}(p)  \rbrace)$ and $\textrm{Pr}( \cap_{\bm{s}_n \in \mathcal{D}_0} \lbrace Y_{t_0}(\bm{s}_n) > Q^{(n)}_{t_0}(p) \rbrace )$ for the same choices of $\mathcal{D}_0$ as above. Let $Q^{(n)}(p)$ be the $p$-th quantile function of the (temporally invariant) distribution of $\varepsilon_{t_0}(\bm{s}_n)$. Following (\ref{eq1}), we have $Q^{(n)}_{t_0}(p) = \mu_{t_0}(\bm{s}_n) + Q^{(n)}(p)$, so ${\textrm{Pr}( \cup_{\bm{s}_n \in \mathcal{D}_0} \lbrace Y_{t_0}(\bm{s}_n) > Q^{(n)}_{t_0}(p)  \rbrace)}=\textrm{Pr}\left( \cup_{\bm{s}_n \in \mathcal{D}_0} \lbrace \varepsilon_{t_0}(\bm{s}_n) > Q^{(n)}(p)  \rbrace \right)$, which does not change over time. Similarly, we have that $\textrm{Pr}( \cap_{\bm{s}_n \in \mathcal{D}_0} \lbrace Y_{t_0}(\bm{s}_n) > Q^{(n)}_{t_0}(p) \rbrace)=\textrm{Pr}\left( \cap_{\bm{s}_n \in \mathcal{D}_0} \lbrace \varepsilon_{t_0}(\bm{s}_n) > Q^{(n)}(p)  \rbrace \right)$. From low to high values of $p$, we present the two types of exceedance probabilities in the bottom panel of Figure \ref{temperature_exceedance}.  At a marginal quantile level of 0.9, the estimated probabilities range within $(0.0088, 0.3605)$, $(0.0070, 0.3933)$ and $(0.0155, 0.3014)$ for the three regions, respectively, indicating high risk of simultaneously large temperatures in these regions.

\subsection{Estimated hotspots}
\label{estimated hotspots}
After exploring the joint threshold exceedance probabilities for fixed regions, we now focus on identifying the regions at risk themselves, i.e., hotspots, for the year 2099. We estimate the $95\%$ confidence regions $\mathcal{D}_{u^+}^0$ for joint exceedance levels $u=33^\circ$C, $u=33.5^\circ$C and $u=34^\circ$C. The maps of $\mathcal{D}_{u^+}^0$ for Weeks 34 and 40 are presented in Figure~\ref{exceedance_regions}. The results for other summer weeks are provided in the Supplementary Material. 

\begin{figure}[t!]
\centering
\adjincludegraphics[height = 0.84\linewidth, width = 0.84\linewidth, trim = {{.17\width} {.33\width} {.17\width} {.33\width}}, clip]{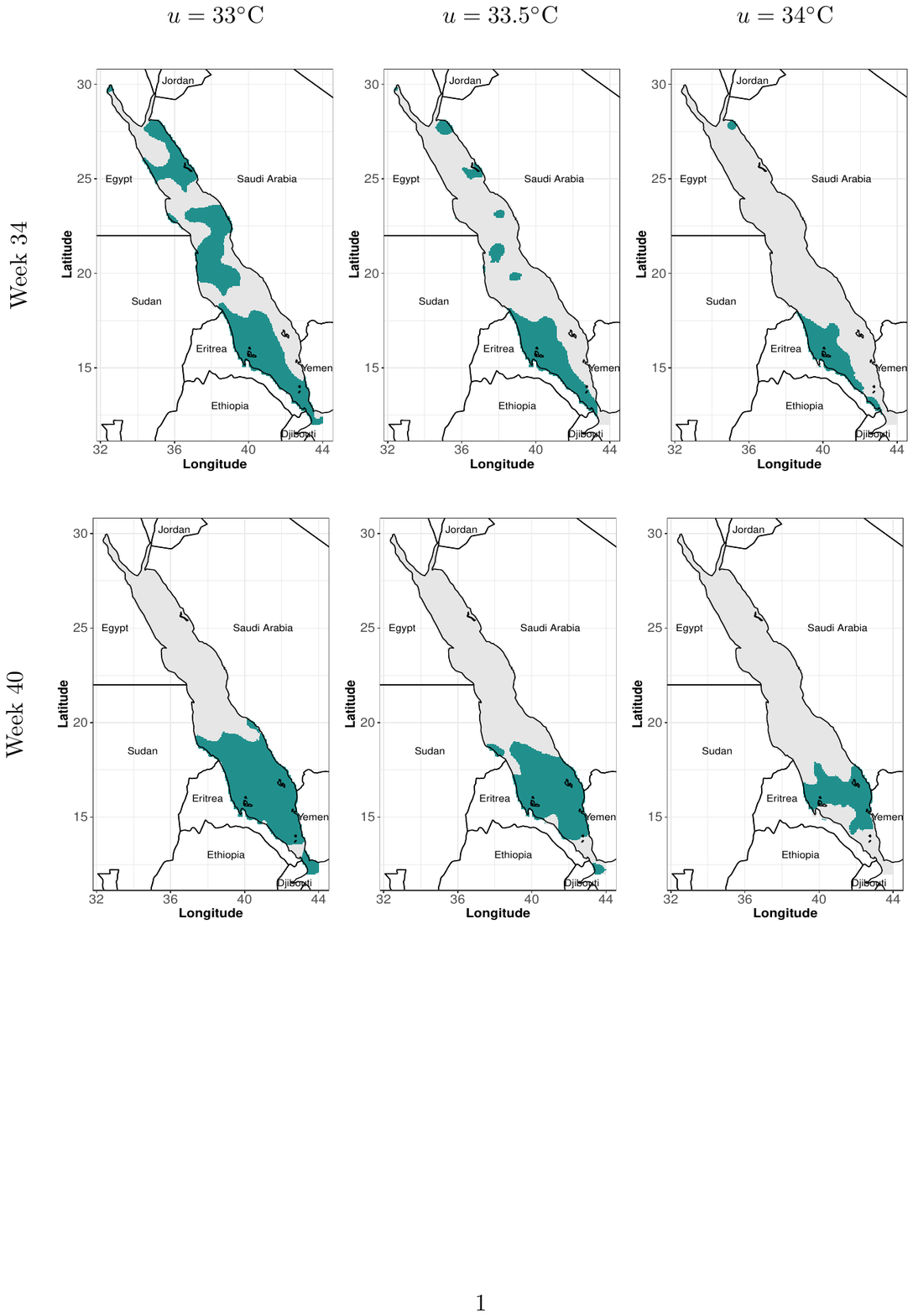}
\caption{The $95\%$ confidence regions $\mathcal{D}_{u^+}^0$ of the Red Sea SST profile, projected to year 2099 (corresponds to the highest RCP 8.5-based SST projection until 2100), for exceedance levels $u=33^\circ$C, $u=33.5^\circ$C and $u=34^\circ$C for Weeks 34 and 40.}
\label{exceedance_regions}
\end{figure}

For Week 40, $\mathcal{D}_{u^+}^0$ stretches over a major portion of the southern Red Sea for each temperature threshold considered. When $u=33^\circ$C, the estimated $\mathcal{D}_{u^+}^0$ covers almost entirely the area within the latitudes 13.5$^\circ$N and 19$^\circ$N and also a small region near coastal Djibouti. When $u=33.5^\circ$C, the estimated $\mathcal{D}_{u^+}^0$ covers a large region between the latitudes 14$^\circ$N and 18$^\circ$N. Finally, when $u=34^\circ$C, the estimated $\mathcal{D}_{u^+}^0$ covers a narrower region stretched between the latitudes 15.5$^\circ$N and 17.5$^\circ$N, including in particular the Dahlak Islands of Eritrea and Farasan Islands of Saudi Arabia, which host many endemic species including corals. The number of grid cells within $\mathcal{D}_{u^+}^0$ are 7136, 5487, and 2819 for the three different thresholds, respectively. For Week 34, the estimated hotspots include 8905, 4186, and 2321 grid cells for the three different thresholds, respectively, and mostly affect the South-Western part of the Red Sea for high SST thresholds.

{While the estimated hotspots based on RCP 8.5 presented here are quite alarming, the results for RCP 4.5 (see the Supplementary Material) are slightly less concerning. Nevertheless, from our analysis, it is likely that large parts of the southern Red Sea including major coral reefs will experience joint extreme SST events within the current century, which may have important consequences in terms of coral bleaching and mortality. As corals are believed to be somewhat resilient to their local environment, it would also be interesting from an applied perspective to estimate hotspots for spatially-varying thresholds. Moreover, for a better risk assessment of coral bleaching, a future research direction could be to extend our model for capturing spatiotemporal dependence, in order to measure the persistence of high SST values over time, and eventually estimate spatiotemporal hotspots.}






\section{Discussion and perspectives}
\label{discussions}

In this paper, we have proposed a low-rank semiparametric Bayesian spatial model for high-dimensional spatiotemporal data with spatial tail-dependence, where the observations are assumed to be independent across time. The proposed model has a flexible mean structure that captures trend, seasonality and accounts for spatial variability in the mean component. Using B-splines for modeling seasonality and spatial variability helps to identify local spatiotemporal features. Relaxing the parametric Gaussian process (GP) assumption that is generally used in the analysis of high-dimensional spatial data, we propose here a flexible semiparametric model that better captures the marginal and dependence structures. While a finite Dirichlet process mixture of GPs relaxes the parametric GP assumption, it is not apt for modeling spatial data with strong extremal dependence and we circumvent this issue through a mixture of low-rank Student's $t$ processes. Furthermore, the covariance structure of the proposed model is constructed from empirical orthogonal functions (EOFs) and it provides a reasonable sparse approximation to the highly nonstationary sample spatial covariance, while allowing for inference in high dimensions.  We have also developed a hotspot estimation method tailored for our proposed model, which allows us to identify regions at risk of joint extreme events.


{ Our statistical analysis revealed several important features of the Red Sea SST data. The decadal rate of change in mean SST varies spatially as well as seasonally; during winter weeks, the highest values are near the latitude 22$^\circ$N while the lowest values are observed near the southern end of the Red Sea. During summer weeks, while the mean SST is generally lower in the northern Red Sea (particularly, near the Gulf of Aqaba and Suez) compared to other regions, its increasing rate is higher in those regions compared to the southern part. Considering the overall decadal rate of change (averaged across weeks), the estimates are positive and significant at every grid cell; higher values are observed all along the coast of the northern Red Sea. Moreover, when including RCP 8.5-based SST as the covariate, the estimated 50-year return levels turn out to be very high ($\approx 35$--$36^\circ$C) during the warmest weeks in the southern Red Sea across a large region stretching from Eritrea to South-Western Saudi Arabia. Furthermore, joint exceedance probabilities for regions near the Dahlak Islands and the Farasan Islands are estimated to be alarmingly high. Both areas host major coral reefs and our study indicates a high chance that the sea surface temperature will jointly exceed high thresholds, which has a direct impact on the risk of coral bleaching (even though several other factors than SST are also important). Our estimation and prediction of hotspots is useful to identify regions ``at risk'' from an environmental and ecological perspective. Considering the very high (fixed) threshold of $34^\circ$C, the estimated hotspots based on the RCP 8.5 scenario indeed cover the Dahlak Islands and Farasan Islands. Climate change mitigation measures might be necessary to safeguard coral reefs and ecosystems, which are not only important for ecological reasons but also because they support tourism in these regions.}


While the proposed model is able to capture several features of the data very flexibly, it has a few downsides. A first drawback is that the model assumes that the temporal replicates are independent in time, which may not be realistic with daily or sub-daily data. The spatiotemporal model of \cite{hazra2018semiparametric}, which assumes a copula structure in time, is a possible solution to address this issue, but it is limited to small spatiotemporal datasets. No closed form expression of the full posterior exists for the random effects involved in that model and hence the analysis becomes computationally challenging for large temporal dimensions. The MCMC based on deterministic transformations proposed by \cite{dutta2014markov} might be a computationally feasible solution. Additionally, the model of \cite{hazra2018semiparametric} assumes that the temporal dependence structure is spatially invariant which may not be a realistic assumption for large geographic regions like the Red Sea. Another drawback is that our proposed model does not capture independence as the distance between sites increases, and its spatial extremal dependence remains nonzero throughout the entire spatial domain. Although this does not seem to be an issue for our Red Sea data, which are strongly dependent, it could be problematic in other data applications with shorter-range spatial dependence. The approach of \cite{morris2017space}, which breaks down long-range dependence by introducing random partitions of the spatial domain, could be a possible solution to address this issue.

\bibliographystyle{apalike}

\baselineskip=16pt

\bibliography{spatialextremes}
\end{document}